\documentclass[noinfoline]{imsart}

\usepackage[a4paper,top=2.5cm, bottom=3cm, left=2.5cm, right=2.5cm]{geometry}
\usepackage{fancyhdr,ulem}

\RequirePackage{amsthm,amsmath,amsfonts,amssymb}

\allowdisplaybreaks
\RequirePackage[colorlinks,citecolor=blue,urlcolor=blue]{hyperref}
\usepackage[capitalize]{cleveref}
\usepackage{mathrsfs} 

\usepackage{graphicx}
\graphicspath{{Images/}}
\usepackage{float}
\usepackage[figuresright]{rotating} 
\usepackage{pdfpages}
\usepackage{wrapfig}
\usepackage{caption}

\usepackage{enumerate}
\usepackage{enumitem} 
\usepackage{tikz}
\usepackage{algorithm}
\usepackage[noend]{algpseudocode}

\makeatletter
\def\algbackskip{\hskip-\ALG@thistlm}
\makeatother

\renewcommand{\b}[1]{\mathbb{#1}} 
\newcommand{\m}[1]{\mathbf{#1}} 
\newcommand{\s}[1]{\mathscr{#1}} 
\renewcommand{\c}[1]{\mathcal{#1}} 

\newcommand{\vertii}[1]{\left\lVert #1 \right\rVert}
\newcommand{\innpr}[2]{{\left\langle {#1},{#2}\right\rangle}} 

\newcommand{\rk}{\mathrm{k}} 

\DeclareMathOperator*{\DICT}{dict} 
\DeclareMathOperator*{\RKHS}{RK} 

\newcommand{\Cov}{\text{Cov}} 

\def\multiset#1#2{\ensuremath{\left(\kern-.3em\left(\genfrac{}{}{0pt}{}{#1}{#2}\right)\kern-.3em\right)}}

\DeclareMathOperator*{\tr}{tr} 
\DeclareMathOperator*{\ovec}{vec}

\DeclareMathOperator*{\odiag}{diag}
\DeclareMathOperator*{\spann}{span}

\theoremstyle{plain}
\newtheorem{theorem}{Theorem}[section]

\newtheorem{proposition}[theorem]{Proposition}
\newtheorem{definition}[theorem]{Definition}
\newtheorem{remark}[theorem]{Remark}
\newtheorem*{remark*}{Remark}

\begin{document}

\pagestyle{fancy}
\fancyhead{}
\renewcommand{\headrulewidth}{0pt} 
\fancyhead[CE]{H. Yun \& V.~M. Panaretos}
\fancyhead[CO]{Fast and Cheap Covariance Smoothing}

\begin{frontmatter}
\title{{\large Fast and Cheap Krylov-Based Covariance Smoothing}}

\begin{aug}
\author[A]{\fnms{Ho} \snm{Yun}\ead[label=e1]{ho.yun@epfl.ch}}
\and
\author[A]{\fnms{Victor M.} \snm{Panaretos}\ead[label=e2,mark]{victor.panaretos@epfl.ch}}

\thankstext{t1}{Research supported by a Swiss National Science Foundation grant.}
\address[A]{Ecole Polytechnique F\'ed\'erale de Lausanne, \printead{e1,e2}}
\end{aug}

\begin{abstract}
We introduce the Tensorized-and-Restricted Krylov (TReK) method, a simple and efficient algorithm for estimating covariance tensors with large observational sizes. TReK extends the Krylov subspace method to incorporate range restrictions, enabling its use in a variety of covariance smoothing applications. By leveraging tensor-matrix operations, it achieves significant improvements in both computational speed and memory cost, improving over existing methods by an order of magnitude. TReK ensures finite-step convergence in the absence of rounding errors and converges fast in practice, making it well-suited for large-scale problems. The algorithm is element-free and highly flexible, supporting a wide range of forward and projection tensors.
\end{abstract}


\end{frontmatter}

\maketitle


\section{Introduction}
\begin{figure}[h!]
\centering
\includegraphics[width=.8\textwidth]{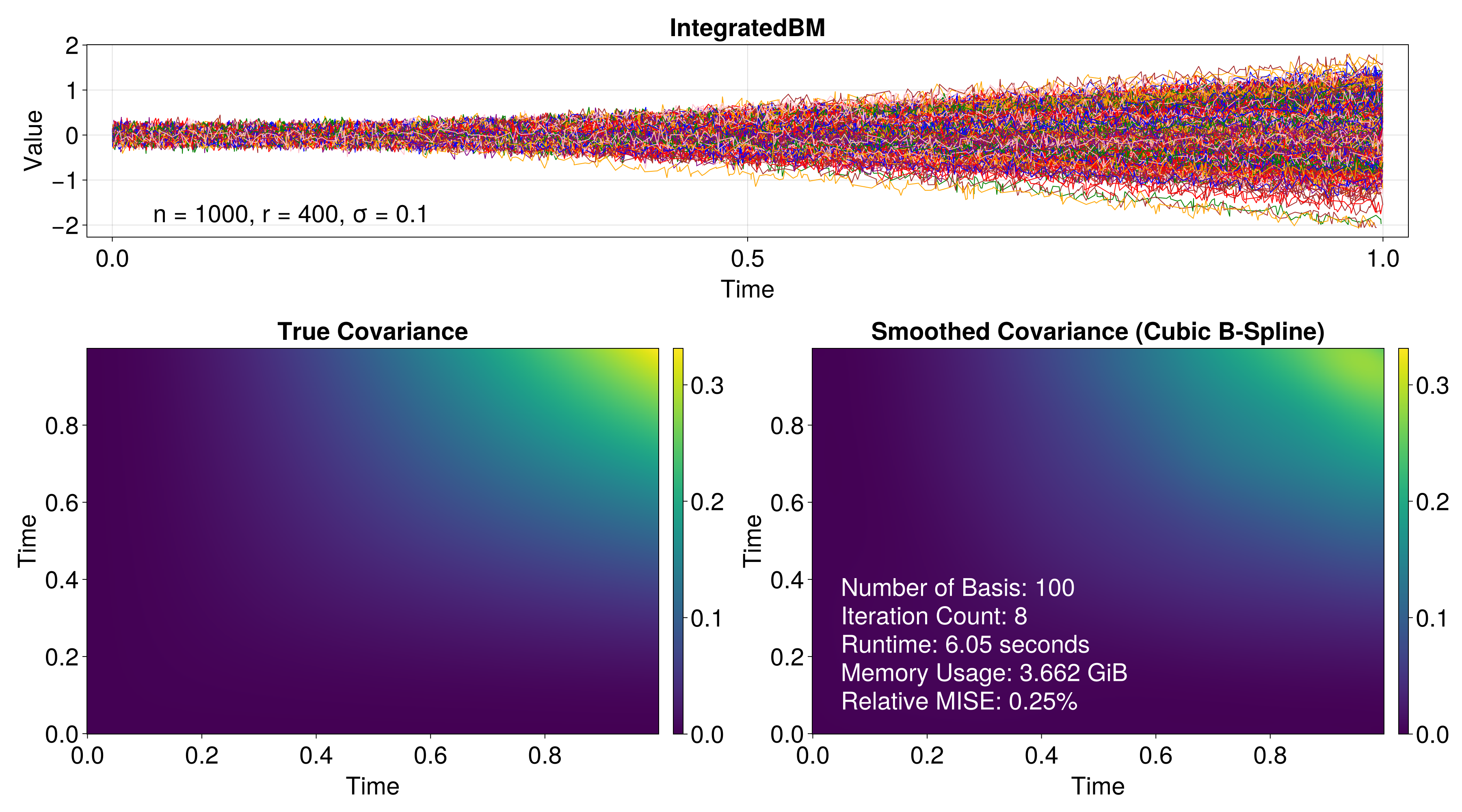}
\caption{\textbf{Top :} $n=1000$ sample paths of the Integrated Brownian motion, each randomly observed at $r=400$ points, with added noise at a level of $\sigma=0.1$. \textbf{Bottom :} true covariance function (left) and its estimator (right) using 100 bivariate cubic B-splines. The estimator reached the stopping criterion after 8 iterations in 6.05 seconds with a memory footprint of 3.66 [GB] on 6 threads of Apple M1 Pro. In contrast, direct solvers without sparsification require at least 120 [GB] of storage merely to form \(\m{F}_{\otimes}\) explicitly.}
\label{fig:BM:Intro}
\end{figure}

This work introduces the Tensorized-and-Restricted Krylov (\texttt{TReK}) method for scalable covariance smoothing in Functional Data Analysis (FDA). 
Rather than propose a new estimation scheme, we develop a class of algorithms that remain computationally efficient at problem scales where conventional numerical approaches become impractical. Our algorithm can be used with both commonly used ``pool-then-smooth" estimation schemes, namely spline-based and RKHS-based methods. Underlying our proposal is a careful analysis of the algebraic structure of the problem. Our contributions are threefold:
\begin{itemize}[leftmargin = *]
    \item A unified block tensor-matrix formulation of least-squares problems in various pool-then-smooth covariance estimation methods,
    \item An efficient implementation of Krylov subspace methods tailored to this formulation,
    \item An adaptation of the Lanczos algorithm for Functional Principal Component Analysis (FPCA).
\end{itemize}
Our methodology is both theoretically grounded and practically versatile: it supports sparse or dense designs, regular or irregular sampling schemes, and --as mentioned-- allows for estimation using either B-splines or reproducing kernels. In sparse settings, \texttt{TReK} yields near-instantaneous solutions (fast); in large-scale problems where existing methods struggle, it remains computationally tractable (cheap) as shown in \cref{fig:BM:Intro}. While our focus is on FDA, the core principles of \texttt{TReK} are more broadly applicable to problems where multivariate interactions and structural constraints must be preserved, such as in multi-dimensional ANOVA models \cite{currie2006generalized, gu1993semiparametric}, digital antenna array design \cite{slyusar1998model, slyusar1999family, sidiropoulos2002khatri}, space-time MIMO communication systems \cite{de2013space, ximenes2015semi, batselier2017tensor}, and heterogeneous tomography \cite{yun2025computerized, yun2025low}.

In FDA, the goal (or at least starting point) is to estimate  the mean or covariance function of i.i.d. random functions, which are observed discretely with noise. Mean estimation using $P$ basis functions from $N$ observations leads to a standard \emph{matrix-vector} least-squares problem:
\begin{equation*}
    \min_{\m{a}} \|\m{F} \m{a} - \m{y} \|^{2} \quad \Longleftrightarrow \quad  (\m{F}^{\top} \m{F}) \m{a} = \m{F}^{\top} \m{y},
\end{equation*}
where $\m{F} \in \b{R}^{N \times P}$ is the forward matrix, $\m{y} \in \b{R}^{N}$ is the observed data vector, and $\m{a} \in \b{R}^{P}$ contains the coefficients to be estimated in the chosen basis. Crucially, the formulation should be invariant under reindexing or basis changes -- a property that holds mathematically, but not computationally when using direct factorization-based solvers (e.g., SVD, Cholesky or LDL\textsuperscript{T}), which could be sensitive to sparsity patterns. Even minor permutations can introduce significant fill-in and degrade numerical stability, thus necessitate pivoting \cite{nocedal1999numerical, hansen1998rank, golub2013matrix}. 


Covariance estimation via pooling presents additional complexity. The target function has two arguments, so the resulting regression problem now becomes a \emph{tensor–matrix} equation:
\begin{equation}\label{eq:intro:cov:tens}
    \min_{\m{A}} \|\s{F} \m{A} - \m{Y} \|^{2} \quad \Longleftrightarrow \quad  (\s{F}^{*} \s{F}) \m{A} = \s{F}^{*} \m{Y},
\end{equation}
where $\s{F}$ is a structured forward operator incorporating:  
\begin{itemize}[leftmargin = *]
    \item a block-wise tensor product (Khatri–Rao product) \cite{khatri1968solutions, neudecker1995hadamard, liu2008hadamard} of the mean-design matrix $\m{F}$ that encapsulates the second-order dependence;
    \item an elimination operator \cite{magnus1980elimination} that discards diagonal entries of $\m{Y}$ contaminated by additive noise \cite{staniswalis1998nonparametric, yao2005functional}.
\end{itemize}

A common strategy in the literature is to flatten the block tensor-matrix equation \eqref{eq:intro:cov:tens} via vectorization, reducing it to a matrix–vector problem:
\begin{equation}\label{eq:intro:cov:flat}
    \min_{\m{a}_{\otimes}} \|\m{F}_{\otimes} \m{a}_{\otimes} - \m{y}_{\otimes} \|^{2} \quad \Longleftrightarrow \quad  (\m{F}_{\otimes}^{\top} \m{F}_{\otimes}) \m{a}_{\otimes} = \m{F}_{\otimes}^{\top} \m{y}_{\otimes},
\end{equation}
after which direct solvers are applied \cite{wood2008fast, xiao2018fast, gu1993semiparametric, greven2011longitudinal, cederbaum2018fast, eilers2003multivariate, wood2017p, eilers1996flexible, kauermann2011functional, xiao2016fast, li2020fast}. However, once the operator $\s{F}$ is explicitly represented as the matrix $\m{F}_{\otimes}$, its underlying algebraic structure is lost when handed off to generic direct solvers. More critically, this approach demands the explicit construction of either $\m{F}_{\otimes}$, its Gram matrix $\m{F}_{\otimes}^{\top} \m{F}_{\otimes}$, or its pseudo inverse $(\m{F}_{\otimes})^{\dagger}$, which can incur considerable computational and memory overhead \cite{hanke2001lanczos, engl1996regularization, hanke2017conjugate, golub2013matrix, chung2024computational, chung2008weighted}. 
These challenges have popularized methods based on B-splines, where a tall-and-skinny sparse matrix $\m{F}_{\otimes}$ can be efficiently evaluated through the de-Boor recursion \cite{de1972calculating, de1978practical, cox1981practical, eilers1996flexible}. While efficient, such approaches lack adaptability to the data-driven geometries, including irregular grids or manifold domains. By contrast, RKHS-based methods \cite{cai2010nonparametric, caponera2022functional, cai2011optimal, yun2025computerized, stoecker2024functional} offer greater data-adaptability but rely on  the dense semi-positive definite (s.p.d.) kernel Gram matrix $\m{F}_{\otimes}$, which scales poorly with increasing data size and thus require aggressive truncation.

To address these limitations, we propose solving the system iteratively using Krylov subspace methods (e.g., CGLS, GMRES, MINRES, LSQR, or LSMR) \cite{saad2003iterative, hestenes1952methods, greenbaum1997iterative, bjorck2024numerical, paige1975solution, paige1982lsqr, paige1982algorithm, fong2011lsmr}, selected based on the properties of the forward matrix $\m{F}_{\otimes}$. These methods work by projecting the solution onto a growing sequence of low-dimensional Krylov subspaces, generated by only one matrix–vector product per iteration with $\m{F}_{\otimes}$ (and with $\m{F}_{\otimes}^{\top}$ in case of assymetry), unlike SVD-based methods \cite{lanczos1950iteration, golub2013matrix, arnoldi1951principle}. Consequently, these methods are inherently \emph{element-free}: they rely only on the action of $\m{F}_{\otimes}$ rather than its explicit components, and are thus well-suited for large-scale problems in FDA as vectorization tricks can be applied to its action. Furthermore, unlike conventional approaches that rely on flattening the problem as in \eqref{eq:intro:cov:flat}, Krylov methods can operate directly on the native inner-product structure of \eqref{eq:intro:cov:tens}, provided this structure is preserved via an inner-product isomorphism. This raises a key question: \emph{what is the natural structure governing the action of $\m{F}$, and how can it be exploited to avoid redundant computation?}

Given that data pairs within each random function reflect the symmetric nature of covariance functions, we model the data $\m{Y}$ as a symmetric block diagonal matrix. Instead of half-vectorization \cite{cederbaum2018fast}, we apply the operator directly to these symmetric matrix blocks, excluding noise-contaminated entries. This yields a \emph{block tensor–matrix} equation that can be solved iteratively -- avoiding the boilerplate code for vectorization and streamlining the FPCA pipeline through the Lanczos algorithm \cite{lanczos1950iteration, wang2024adaptive}. Compared to existing approaches, our method naturally promotes sparsity, avoids \textit{ad hoc} flattening schemes, and adheres closely to theoretical insights. Consequently, it is readily extensible to broader applications, including signal processing \cite{slyusar1998model, slyusar1999family, sidiropoulos2002khatri, de2013space, ximenes2015semi, batselier2017tensor} and functional inverse problems  \cite{hanke2017taste, yun2025computerized, yun2025low}.

Whether using the spline or RKHS estimation scheme, regularization is essential -- both from a theoretical and practical standpoint -- to prevent overfitting and ensure numerical stability \cite{hanke2001lanczos, engl1996regularization, hanke2017conjugate}. While Tikhonov regularization is commonly employed in direct methods for its computational convenience, it still becomes impractical in large-scale problems due to the costly process of tuning the penalty parameter \cite{chung2024computational}. In contrast, the Krylov methods we propose provide implicit regularization via early stopping (still allowing additional penalty functionals to be incorporated, if desired). The iteration count itself serves as a surrogate regularization parameter, balancing data fidelity and numerical stability by controlling how closely the projected subproblem approximates the original ill-posed system \cite{hanke2001lanczos, engl1996regularization, hanke2017conjugate}.

We refer to this conceptual class of structured, element-free approaches as the \texttt{TReK} method. By circumventing explicit construction of large-scale operators, \texttt{TReK} offers substantial gains in both speed and memory usage. This dramatic reduction highlights \texttt{TReK}'s potential to efficiently process large datasets, making it a scalable and economical alternative to conventional covariance smoothers.

\section{Notation and Background}
Throughout the paper, \emph{bold uppercase} symbols denote matrices, while \emph{bold lowercase} symbols represent vectors. For an s.p.d. matrix $\m{P}$, we denote the weighted $\m{P}$-inner product by $\innpr{\m{c}}{\m{d}}_{\m{P}} := \m{c}^{\top} \m{P} \m{d}$, and the $\m{P}$-norm by $\|\m{c}\|_{\m{P}} := (\m{c}^{\top} \m{P} \m{c})^{1/2}$, which is a semi-norm in case of rank-deficiency. \emph{Script-style} symbols refer to linear operators (e.g., $\s{F}$) acting on matrices, and we denote the corresponding Moore–Penrose pseudoinverse by $\s{F}^{\dagger}$, and the range by $\c{R}(\s{F})$ \cite{golub2013matrix, hsing2015theoretical, hanke2017taste}. \emph{Calligraphic-style} symbols denote spaces or sets. Scalar quantities are written in standard (non-bold) font, with key dimensions as follows:
\begin{itemize}
    \item $n$: the number of random functions,
    \item $r_{i}$: the number of measurement locations for each random function,
    \item $g$: the number of grid points to evaluate the mean function; the covariance function is evaluated over a $g \times g$ grid.
\end{itemize} 

\begin{definition}
Let $\b{H}$ be a Hilbert space of real-valued functions defined on a set $\c{T}$. A bivariate function $K:\c{T} \times \c{T}\rightarrow \b{R}$ is called a reproducing kernel for $\b{H}$ if
\begin{enumerate}
    \item For any $t \in \c{T}$, a feature map $\rk_{t}(\cdot):= K(\cdot, t)$ at $t \in \Omega$ belongs to $\b{H}$.
    \item For any $f \in \b{H}$, the point evaluation at $t \in \c{T}$ is given by $f(t)=\langle f, \rk_{t} \rangle$.
\end{enumerate}
A Hilbert space equipped with a reproducing kernel is called a Reproducing Kernel Hilbert Space (RKHS). 
\end{definition}
By the Moore–Aronszajn theorem \cite{aronszajn1950theory}, any reproducing kernel is an s.p.d. function. Conversely, any s.p.d. function induces a unique RKHS, justifying the notation $\b{H}=\b{H}(K)$ \cite{paulsen2016introduction,hsing2015theoretical}. When the elementary tensor between $f_{1}, f_{2} \in \b{H}$ is understood as a bivariate function, the reproducing property yields
\begin{equation}\label{eq:tens:repro:prop}
    (f_{1} \otimes f_{2})(t_{1}, t_{2}) = \innpr{(f_{1} \otimes f_{2})}{(\rk_{t_{1}} \otimes \rk_{t_{2}})} = \innpr{f_{1}}{\rk_{t_{1}}} \innpr{f_{2}}{\rk_{t_{2}}} = f_{1}(t_{1}) f_{2}(t_{2}).
\end{equation}

\subsection{Data Setup}
Consider a sample of i.i.d. second-order random functions $X_{1}, \cdots, X_{n}$ on $\c{T}$, where $\b{E}[X_{i}(t)^{2}] < \infty$ for $i=1, \dots, n$ and $t \in \c{T}$, that are mean-square continuous and jointly measurable \cite{hsing2015theoretical, kallenberg1997foundations}. Let $\mu: \c{T} \to \mathbb{R}$ denote the mean function, and $\Gamma, \Sigma: \c{T} \times \c{T} \to \mathbb{R}$ the second moment and covariance functions, respectively: 
\begin{equation}\label{eq:def:moments}
    \mu(t) = \b{E}[X_{1}(t)], \quad \Gamma(t_{1}, t_{2}) = \b{E}[X_{1}(t_{1}) X_{1}(t_{2})], \quad \Sigma(t_{1}, t_{2}) = \Gamma(t_{1}, t_{2}) - \mu(t_{1}) \mu(t_{2}), \quad t, t_{1}, t_{2} \in \c{T}.
\end{equation}
In practice, these random functions are observed at (random) locations $t_{ij} \in \c{T}$, subject to i.i.d. noise $\varepsilon_{ij} \in \b{R}$ with variance $0<\sigma^{2} <\infty$, 
\begin{equation*}
    y_{ij} = X_{i} (t_{ij}) + \varepsilon_{ij}, \quad 1 \le i \le n, 1 \le j \le r_{i},
\end{equation*}
where $X_{i}$, $t_{ij}$, and $\varepsilon_{ij}$ are mutually independent. 
To estimate the mean function $\mu$, we minimize the following \emph{unregularized} empirical risk functional:
\begin{equation}\label{eq:loss:1st:intro}
    \c{L}(\mu) = \sum_{i=1}^{n} \sum_{j=1}^{r_{i}} (y_{ij} - \mu(t_{ij}))^{2},
\end{equation}
which pools all observed data, thus effectively treating the observations as samples from a single process. 

For covariance estimation, the structure is more nuanced. While cross-subject pairs ($i_1 \ne i_2$) are uninformative due to independence, within-subject \emph{off-diagonal} pairs
\begin{equation*}
    \c{O}_{i} := \{(j_{1}, j_{2}): 1 \le j_{1} \neq j_{2} \le r_{i}\}, \quad 1 \le i \le n,
\end{equation*}
capture the second-order structure \cite{staniswalis1998nonparametric, yao2005functional}. This is because the conditional covariance of any two observations satisfies: 
\begin{align*}
    \Cov[y_{i_{1}j_{1}} , y_{i_{2}j_{2}} \vert t_{i_{1}j_{1}}, t_{i_{2}j_{2}} ] 
    =
    \begin{cases}
        0, &i_{1} \neq i_{2}, \\
        \Sigma(t_{i_{1}j_{1}}, t_{i_{2}j_{2}}) + \sigma^{2} \delta_{j_{1}j_{2}}, &i_{1} = i_{2}
    \end{cases},
\end{align*}
which motivates the exclusion of diagonal products to estimate $\Gamma$ and  obtain a plug-in estimator for $\Sigma$ as $\hat{\Sigma} = \hat{\Gamma} - \hat{\mu} \otimes \hat{\mu}$ \cite{caponera2022functional}. 
To reformulate this as a least-square problem, we define the elimination operator with respect to $\c{O}_{i}$: 
\begin{equation*}
    \s{O}_{i}: \b{R}^{r_{i} \times r_{i}} \to \b{R}^{r_{i} \times r_{i}}, \quad (\s{O}_{i} \m{C}_{i})[j_{1}, j_{2}] =
    \begin{cases}
        \m{C}_{i}[j_{1}, j_{2}] &, \quad (j_{1}, j_{2}) \in \c{O}_{i} \\
        0 &, \quad (j_{1}, j_{2}) \notin \c{O}_{i}
    \end{cases}, \qquad 1 \le i \le n.
\end{equation*}
Consequently, the associated \emph{unregularized} loss becomes:
\begin{align}\label{eq:loss:2nd:intro}
    \c{L}_{\otimes}(\Gamma)
    = \sum_{i=1}^{n} \sum_{(j_{1}, j_{2}) \in \c{O}_{i}} \left(y_{ij_{1}} y_{ij_{2}} -  \Gamma(t_{ij_{1}}, t_{ij_{2}}) \right)^{2}. 
\end{align}

\begin{remark}\label{rmk:centred}
We remark that several works \cite{cai2010nonparametric, xiao2013fast, xiao2018fast, cederbaum2018fast, li2020fast} have proposed directly estimating the covariance $\Sigma$ using \emph{centered} observations $\tilde{y}_{ij}:= y_{ij} - \hat{\mu}(t_{ij})$, which yields the following empirical risk functional:
\begin{align*}
    \tilde{\c{L}}_{\otimes}(\Sigma)
    &:= \sum_{i=1}^{n} \sum_{(j_{1}, j_{2}) \in \c{O}_{i}} (\tilde{y}_{ij_{1}} \tilde{y}_{ij_{2}} - \Sigma(t_{ij_{1}}, t_{ij_{2}}) )^{2},
\end{align*}
when the penalty functional is disregarded, and the noise level $\sigma^{2}$ is known or treated as a nuisance parameter in their formulations. Whether we estimate $\Gamma$ first and subsequently compute $\hat{\Sigma} = \hat{\Gamma} - \hat{\mu} \otimes \hat{\mu}$, or instead incorporate $\hat{\mu}$ into the risk functional, the resulting linear equation remains effectively the same. Consequently, the numerical complexity remains unchanged, so we focus on the plug-in approach via \eqref{eq:loss:1st:intro} and \eqref{eq:loss:2nd:intro}.
\end{remark}

\subsection{Least-Squares Problem}
Technical derivations in this section are deferred to \cref{sec:cov:est,sec:cplxty}. To enable a unified comparison across different estimation methods, we adopt a generic notation: $\c{D}$, $\c{S}$, and $\c{C}$ respectively denote the data space, the search space, and the coefficient space. Their dimensions are denoted by $N = \dim(\c{D})$ and $P = \dim(\c{S}) = \dim(\c{C})$.
When explicitly distinguishing between first- and second-order estimation problems, we use a subscript $\otimes$ to indicate quantities associated with covariance estimation (e.g., $\c{D}_{\otimes}, N_{\otimes}$), in contrast to their mean estimation counterparts (e.g., $\c{D}, N$). 
The underlying estimation framework -- such as dictionary-based or RKHS-based -- is indicated via superscripts (e.g., $\c{C}_{\otimes}^{\DICT}$ vs.\ $\c{C}_{\otimes}^{\RKHS}$). 

Let $\m{r} := [r_{1}, \dots, r_{n}] \in \b{N}^{n}$ denote a sequence of the number of observed locations for $i$-th sample path, and denote $|\m{r}| := \sum_{i=1}^{n} r_{i}$, $|\m{r}^{2}| := \sum_{i=1}^{n} r_{i}^{2}$. By the Cauchy-Schwarz inequality, we have $n |\m{r}^{2}| \ge |\m{r}|^{2}$ with equality holds if and only if $r_{i} \equiv r$ so that $|\m{r}| = nr$ and $|\m{r}^{2}| = nr^{2}$. 
For notational consistency with the existing literature in this section, we adopt a flattened matrix-vector representation of the second-order equations as in \eqref{eq:intro:cov:flat}, although this is by \emph{no} means how our iterative algorithm is implemented. Accordingly, the elimination operator is expressed in its matrix form 
\begin{equation*}
    \m{O}^{\flat} := \odiag[\m{O}_{i}^{\flat}] = 
    \begin{pmatrix}
    \m{O}_{1}^{\flat} & \cdots & \m{0} \\
    \vdots & \ddots & \vdots \\
    \m{0} & \cdots & \m{O}_{n}^{\flat}
    \end{pmatrix} \in \b{R}^{|\m{r}^{2}| \times |\m{r}^{2}|}, \quad \m{O}_{i}^{\flat} \in \b{R}^{r_{i}^{2} \times r_{i}^{2}} :  \ovec(\m{C}_{i}) \mapsto \ovec(\s{O}_{i} \m{C}_{i}),
\end{equation*}
where the flat symbol $\flat$ indicates that a matrix level operation has been flattened to the vector level. 

First, we systematically organize the observations $y_{ij}$ and $y_{ij_{1}} y_{ij_{2}}$ in \eqref{eq:loss:1st:intro} and \eqref{eq:loss:2nd:intro} as
\begin{equation}\label{eq:data:intro}
    \m{y} = 
    \begin{pmatrix}
    \m{y}_{1} \\
    \m{y}_{2} \\
    \vdots \\
    \m{y}_{n}
    \end{pmatrix} \in \b{R}^{|\m{r}|} =: \c{D}, \quad 
    \m{y}_{\otimes} =
    \begin{pmatrix}
    \m{y}_{1} \otimes_{\flat} \m{y}_{1} \\
    \m{y}_{2} \otimes_{\flat} \m{y}_{2} \\
    \vdots \\
    \m{y}_{n} \otimes_{\flat} \m{y}_{n}
    \end{pmatrix} = 
    \begin{pmatrix}
    \ovec[\m{y}_{1} \m{y}_{1}^{\top}] \\
    \ovec[\m{y}_{2} \m{y}_{2}^{\top}] \\
    \vdots \\
    \ovec[\m{y}_{n} \m{y}_{n}^{\top}]
    \end{pmatrix} \in \b{R}^{|\m{r}^{2}|} =: \c{D}_{\otimes},
\end{equation}
where $\otimes_{\flat}$ represents the Kronecker product as it flattens the bilinear form $\otimes$, leading to $N = |\m{r}|$ and $N_{\otimes} = |\m{r}^{2}|$. Notably, the data space depends solely on the order of the moment being estimated, not on the estimation methods.

Since $\mu$ and $\Gamma$ reside in infinite-dimensional function spaces, the loss functionals in \eqref{eq:loss:1st:intro} and \eqref{eq:loss:2nd:intro} admit infinitely many minimizers without further structural assumptions. While a comprehensive review of the extensive literature is beyond our scope, a common remedy is to constrain the search space to a finite-dimensional subspace that reflects a desired level of smoothness. Under such a basis expansion, empirical risk minimization reduces to solving a least-squares problem in the corresponding coefficient space. Observe that solving \eqref{eq:loss:2nd:intro} highlights the \emph{pool-then-smooth} strategy, in contrast to the more traditional \emph{smooth-then-pool} approach \cite{hall2006properties, zhang2007statistical, ramsay1998curve}.
Although many alternative methods for covariance smoothing have been proposed -- such as \texttt{FACE}, which relies on Nadaraya–Watson kernel regression \cite{chen2017quantifying, wang2016functional, golovkine2025adaptive}, or \texttt{CovNet}, which leverages neural networks \cite{sarkar2022covnet} -- our focus is on two widely used approaches:
\begin{itemize}[leftmargin=*]
\item \textbf{Dictionary Methods}: The function space is predefined via a fixed basis, most often a spline basis. For the mean, we assume 
\begin{equation*}
    \mu \in \c{S}^{\DICT} = \spann \{\phi_{l}: \c{T} \to \b{R}, 1 \le l \le p \}, \quad \c{C}^{\DICT} = \b{R}^{p}, \quad P^{\DICT} = p.
\end{equation*} 
For second-order estimation, the search space becomes the tensor product:
\begin{equation*}
    \Gamma \in \c{S}_{\otimes}^{\DICT} = \spann \{\phi_{l_{1}} \otimes \phi_{l_{2}}: \c{T} \times \c{T} \to \b{R} : 1 \le l_{1}, l_{2} \le p\}, \quad \c{C}_{\otimes}^{\DICT} = \b{R}^{p^{2}}, \quad P_{\otimes}^{\DICT} = p^{2},
\end{equation*}
where $(\phi \otimes \phi') (t, t') := \phi(t) \phi'(t')$ for $(\phi \otimes \phi') \in \c{L}_2(\c{T}) \otimes \c{L}_2(\c{T}) \cong \c{L}_2(\c{T} \times \c{T})$. This bivariate spline framework underlies the majority of existing covariance smoothers \cite{wood2008fast, xiao2018fast, li2020fast, gu1993semiparametric, greven2011longitudinal, cederbaum2018fast, eilers2003multivariate}. 
While other choices $\phi_l$ exist -- such as known eigenfunctions of the covariance operator \cite{peng2009geometric}, or truncated orthonormal bases in $\c{L}_2(\c{T})$ \cite{yao2005functional, staniswalis1998nonparametric} -- B-splines are most commonly chosen \cite{de1978practical, wood2017p, eilers1996flexible, kauermann2011functional}, due to their compact support, which induces sparsity in the linear system. 

\item \textbf{RKHS Methods}: Alternatively, smoothness can be encoded via an s.p.d. kernel $K$, modeling each $X_i$ as a random element in an RKHS $\b{H}(K)$ \cite{cai2010nonparametric, caponera2022functional, cai2011optimal, yun2025computerized, stoecker2024functional, yun2025low}. The representer theorem \cite{kimeldorf1970correspondence, wahba1981spline} ensures that the minimizer of \eqref{eq:loss:1st:intro} lies in the span of feature maps at observed locations: 
\begin{equation*}
    \hat{\mu} \in \c{S}^{\RKHS} = \spann\{\rk_{ij} := K(\cdot, t_{ij}): 1 \le i \le n, 1 \le j \le r_{i} \}, \quad \c{C}^{\RKHS} = \c{D} = \b{R}^{|\m{r}|}, \quad P= N = |\m{r}|.
\end{equation*}
Unlike dictionary methods, the basis functions $\rk_{ij} \in \b{H}(K)$ are data-adaptive. However, this flexibility comes at the cost of \emph{non-parsimony} -- the number of basis functions equals the total sample size.
For the same reason, the minimizer of \eqref{eq:loss:2nd:intro} lies in the span of tensorized feature maps $\{\rk_{ij_{1}} \otimes \rk_{ij_{2}} : 1 \le i \le n, (j_{1}, j_{2}) \in \c{O}_{i} \}$ between within-subject off-diagonal pairs. An equivalent but more convenient formulation is to consider the larger search space over a block square grid
\begin{equation*}
    \hat{\Gamma} \in \c{S}_{\otimes}^{\RKHS} = \spann \{\rk_{ij_{1}} \otimes \rk_{ij_{2}} : 1 \le i \le n, 1 \le j_{1}, j_{2} \le r_{i} \}, \quad \c{C}_{\otimes}^{\RKHS} = \c{D}_{\otimes} = \b{R}^{|\m{r}^{2}|}, \quad P_{\otimes}^{\RKHS} = N_{\otimes} = |\m{r}^{2}|,
\end{equation*}
and instead impose a constraint on the coefficient vector:
\begin{equation*}
    \hat{\m{a}}_{\otimes}^{\RKHS} =
    \begin{pmatrix}
    \hat{\m{a}}_{1}^{\RKHS} \\
    \vdots \\
    \hat{\m{a}}_{n}^{\RKHS}
    \end{pmatrix} \in \b{R}^{|\m{r}^{2}|}, \quad \hat{\m{a}}_{i}^{\RKHS} \in \c{R}(\m{O}_{i}^{\flat}) \subset \b{R}^{r_{i}^{2}}, \quad 1 \le i \le n.
\end{equation*}
\end{itemize}
We remark that, in both methods, the coefficient matrix $\hat{\m{A}}$ satisfying $\hat{\m{a}}_{\otimes} = \ovec(\hat{\m{A}}) \in \c{C}_{\otimes}$ is symmetric, allowing for further dimensionality reduction via half-vectorization \cite{cederbaum2018fast, xiao2018fast, li2020fast}. 

In each approach, minimizing the loss over the function space $\c{S}$ corresponds to solving a least-squares problem with the corresponding coefficient space $\c{C}$, which takes the form:
\begin{equation*}
    \min_{\m{a} \in \c{C}} \vertii{\m{F} \m{a} - \m{y}}_{\c{D}}^{2}, 
\end{equation*}
where $\m{F}:\c{C} \to \c{D}$ is a fixed forward matrix derived from the loss, $\m{y} \in \c{D}$ is the observed data vector, and $\m{a} \in \c{C}$ is the coefficient vector to solve. These components are summarized in \cref{table:LSE:1st,table:LSE:2nd}, with the proof given in \cref{prop:cplxty}. Notably, unlike the data space $\c{D}$, the coefficient space $\c{C}$ and the corresponding forward matrix $\m{F}$ vary depending on the estimation method. We first review the first-order forward matrix:

\begin{itemize}[leftmargin=*]
\item In dictionary methods, the evaluation matrix is given by
\begin{equation*}
    \m{F}^{\DICT} = \m{\Phi} = \begin{pmatrix}
    \m{\Phi}_{1} \\
    \vdots \\
    \m{\Phi}_{n}
    \end{pmatrix} \in \b{R}^{|\m{r}| \times p}, \quad \m{\Phi}_{i}[j, l] := \phi_{l}(t_{ij}), \quad 1 \le i \le n, \, 1 \le j \le r_{i}, \, 1 \le l \le p,
\end{equation*}
representing evaluations of basis functions at observed locations. Note that this evaluation matrix is sparse for the B-spline, and can be obtained by the de-Boor recursion \cite{de1972calculating}.
Since the function space $\c{S}^{\DICT}$ is predefined, the block structure of $\m{F}^{\DICT}$ is partitioned only \emph{row-wise}.

\item In contrast, RKHS method defines the first-order forward matrix
\begin{equation*}
    \m{F}^{\RKHS} = \m{K} = \begin{pmatrix}
    \m{K}_{11} & \cdots & \m{K}_{1n} \\
    \vdots & \ddots & \vdots \\
    \m{K}_{n1} & \cdots & \m{K}_{nn}
    \end{pmatrix} \in \b{R}^{|\m{r}| \times |\m{r}|}, \quad 
    \m{K}_{i_{1}i_{2}}[j_{1}, j_{2}] := K(t_{i_{1}j_{1}}, t_{i_{2}j_{2}}), \quad \quad 1 \le i_{1}, i_{2} \le n,
\end{equation*}
called the kernel Gram matrix, representing inner products between feature maps. Here, the block partitioning is both row- and column-wise, reflecting the fact that the function space $\c{S}^{\RKHS} = \c{D}$ is selected \textit{a posteriori} and tailored to the data.
\end{itemize}

\begin{table}[h!]
\renewcommand{\arraystretch}{1.5}
\caption{Comparison of least-squares formulations for dictionary- and RKHS-based methods in mean estimation. The sparse pattern and the computational complexity of constructing the forward matrix depends on the method-specific details, such as basis evaluation or kernel function intricacy. Yet, both $\m{\Phi}$ and $\m{K}$ can be evaluated in a parallel manner via $\texttt{SIMD}$ (Single Instruction, Multiple Data) in modern operation systems, meaning that these operations are not expensive.}
\centering
\begin{tabular}{|c||c|c|}
\hline
Method & Dictionary-based & RKHS-based \\
\hline
\hline
Data ($\m{y} \in \c{D}$) & $\m{y} \in \b{R}^{|\m{r}|}$ & $\m{y} \in \b{R}^{|\m{r}|}$  \\
\hline
Coefficients ($\m{a} \in \c{C}$) & $\m{a}^{\DICT} \in \b{R}^{p}$ & $\m{a}^{\RKHS} \in \b{R}^{|\m{r}|}$  \\
\hline
Forward Matrix ($\m{F}$) & $\m{\Phi} \in \b{R}^{|\m{r}| \times p}$  & $\m{K} \in \b{R}^{|\m{r}| \times |\m{r}|}$  \\
\hline
\end{tabular}
\label{table:LSE:1st}
\end{table}

Now, we proceed to the second-order forward matrix. Given a basis for the search space $\c{S}_{\otimes}$, the first-order forward matrix acts to the coefficient \emph{matrix} $\m{A}$ both from the left and right to evaluate the term $\Gamma(t_{ij_{1}}, t_{ij_{2}})$ in the loss functional \eqref{eq:loss:2nd:intro}. When $\m{A}$ is vectorized, this bilinear form can be represented via the Kronecker product $\otimes_{\flat}$. However, the Kronecker product is performed in a block-wise manner, commonly referred to as the Khatri-Rao product \cite{khatri1968solutions, neudecker1995hadamard, liu2008hadamard}, denoted by $\odot_{\flat}$. Subsequently, the orthogonal projection $\m{O}^{\flat}$ is applied to restrict evaluation to the index pairs $\c{O}_{i}$ in \eqref{eq:loss:2nd:intro}. These observations yield the following flattened structure for the second-order forward matrix:

\begin{itemize}[leftmargin=*]
\item In dictionary methods, the second-order forward matrix is given by 
\begin{equation*}
    \m{F}_{\otimes}^{\DICT} = 
    \m{O}^{\flat} (\m{\Phi} \odot_{\flat} \m{\Phi}) =
    \begin{pmatrix}
    \m{O}_{1}^{\flat} (\m{\Phi}_{1} \otimes_{\flat} \m{\Phi}_{1}) \\
    \vdots \\
    \m{O}_{n}^{\flat} (\m{\Phi}_{n} \otimes_{\flat} \m{\Phi}_{n})
    \end{pmatrix} \in \b{R}^{|\m{r}^{2}| \times p^{2}}.
\end{equation*}
Here, there is no partitioning in the column direction. In this case, the row-wise Khatri-Rao product is also referred to the face-splitting product \cite{slyusar1999family, slyusar1998model, sidiropoulos2002khatri, de2013space, ximenes2015semi, batselier2017tensor}, which appears naturally when the search space is predefined.

\item In RKHS methods, The second-order forward matrix applies tensorization in both the row and column directions:
\begin{align*}
    \m{F}_{\otimes}^{\RKHS} = 
    \m{O}^{\flat} (\m{K} \odot_{\flat} \m{K}) \m{O}^{\flat} 
    = 
    \begin{pmatrix}
    \m{O}_{1}^{\flat} (\m{K}_{11} \otimes_{\flat} \m{K}_{11}) \m{O}_{1}^{\flat} & \dots & \m{O}_{1}^{\flat} (\m{K}_{1n} \otimes_{\flat} \m{K}_{1n}) \m{O}_{n}^{\flat} \\
    \vdots & \ddots & \vdots \\
    \m{O}_{n}^{\flat} (\m{K}_{n1} \otimes_{\flat} \m{K}_{n1}) \m{O}_{1}^{\flat} & \dots & \m{O}_{n}^{\flat} (\m{K}_{nn} \otimes_{\flat} \m{K}_{nn}) \m{O}_{n}^{\flat}
    \end{pmatrix} 
    \in \b{R}^{|\m{r}^{2}| \times |\m{r}^{2}|},
\end{align*}
which arises naturally when the search space is data-adaptive. Here, we attach $\m{O}^{\flat}$ on both sides so that the sandwich form $\m{F}_{\otimes}^{\RKHS}$ is s.p.d., which facilitates numerical development. Additionally, this projection increases the sparsity of the system matrix, thereby improving computational efficiency, although negligible.
\end{itemize}

\begin{table}[h!]
\renewcommand{\arraystretch}{1.5}
\caption{Comparison of matrix-vector least-squares formulations for dictionary- and RKHS-based methods in second-order estimation. 
Given the first-order forward matrix $\m{F}$, implementing only the action of the second-order forward matrix $\m{F}_{\otimes}$ preserves the underlying tensor structure and reduces computational cost. In contrast, explicitly constructing $\m{F}_{\otimes}$, thereby loosing its underlying structure, or evaluating full matrix-vector products thereof, substantially increases the complexity.}
\centering
\begin{tabular}{|c||c|c|}
\hline
Method & Dictionary-based & RKHS-based \\
\hline
\hline
Data ($\m{y}_{\otimes} \in \c{D}_{\otimes}$) &  $\m{y}_{\otimes} \in \b{R}^{|\m{r}^{2}|}$ &  $\m{y}_{\otimes} \in \b{R}^{|\m{r}^{2}|}$  \\
\hline
Coefficients ($\m{a}_{\otimes} \in \c{C}_{\otimes}$) &  $\m{a}_{\otimes}^{\DICT} \in \b{R}^{p^{2}}$ &  $\m{a}_{\otimes}^{\RKHS} \in \b{R}^{|\m{r}^{2}|}$  \\
\hline
Forward Matrix ($\m{F}_{\otimes}$) &  $\m{O}^{\flat} (\m{\Phi} \odot_{\flat} \m{\Phi}) \in \b{R}^{|\m{r}^{2}| \times p^{2}}$ & $\m{O}^{\flat} (\m{K} \odot_{\flat} \m{K}) \m{O}^{\flat} \in \b{R}^{|\m{r}^{2}| \times |\m{r}^{2}|}$  \\
\hline
Complexity of Element-free Action & $O(|\m{r}| p^{2} + |\m{r}^{2}|p)$ & $O(|\m{r}| |\m{r}^{2}|)$  \\
\hline
Complexity of Explicit Evaluation & $O(|\m{r}^{2}| p^{2})$ & $O(|\m{r}^{2}|^{2})$  \\
\hline
\end{tabular}
\label{table:LSE:2nd}
\end{table}

\subsection{Tikhonov Regularization}\label{sec:tikho:review}

To avoid overfitting, it is common to introduce an s.p.d. penalty matrix $\m{P}:\c{C} \to \c{C}$, selected based on the desired level of smoothness in estimation:
\begin{itemize}[leftmargin=*]
\item In dictionary methods, suppose we aim to penalize total curvature via the squared $L^2$ norm of the second derivative for mean estimation, i.e., $ \|\mu\|_{\m{P}}^{2} = \int_{\c{T}} |\mu^{(2)}(t)|^{2} dt$. This leads to a penalty matrix:
\begin{equation*}
    \m{P}^{\DICT} \in \b{R}^{p \times p}, \quad \m{P}^{\DICT}[l_{1}, l_{2}] := \int_{\c{T}} \phi_{l_{1}}^{(2)}(t) \phi_{l_{2}}^{(2)}(t) dt, \quad 1 \le l_{l}, l_{2} \le p.
\end{equation*}
The term P-spline is often abused to refer to the penalized B-spline \cite{eilers1996flexible, eilers2003multivariate}, where this integral is discretized in practice to yield a penalty of the form
\begin{equation*}
    \m{P}^{\DICT} = (\m{D}^{\DICT})^{\top} \m{D}^{\DICT}, \quad \m{D}^{\DICT} \in \b{R}^{(p-2) \times p},
\end{equation*}
with $\m{D}^{\DICT}$ being a sparse matrix encoding second-order finite differences of the coefficients. Higher-order differencing schemes are also possible but are beyond the scope of this paper.

For second-order estimation, we consider penalizing smoothness using the Laplacian $\Delta = \partial_{1}^{2} + \partial_{2}^{2}$:
\begin{equation*}
    \|\Gamma\|_{\m{P}_{\otimes}}^{2} = \frac{1}{2} \int_{\c{T} \times \c{T}} |\partial_{1} \Gamma(t_{1}, t_{2})|^{2} + |\partial_{2} \Gamma(t_{1}, t_{2})|^{2} dt_{1} dt_{2}.
\end{equation*}
Unlike the thin-plate spline penalty involving $(\partial_{1} + \partial_{2})^{2}$ \cite{duchon1977splines}, the absence of  a cross-term allows a clean Kronecker-structured discretization:
\begin{align*}
    \m{P}_{\otimes}^{\DICT} &= \frac{(\m{P}^{\DICT} \otimes_{\flat} \m{I} + \m{I} \otimes_{\flat} \m{P}^{\DICT})}{2}
    = \frac{(\m{D}_{\otimes}^{\DICT})^{\top} \m{D}_{\otimes}^{\DICT}}{2} 
    \in \b{R}^{p^{2} \times p^{2}},
\end{align*}
where
\begin{align*}
     \m{D}_{\otimes}^{\DICT} = \begin{pmatrix}
        \m{D}^{\DICT} \otimes_{\flat} \m{I} \\
        \m{I} \otimes_{\flat} \m{D}^{\DICT}
    \end{pmatrix} \in \b{R}^{2(p-2)p \times p^2}.
\end{align*}

\item In RKHS methods, the natural regularization is induced by the RKHS norm. For mean estimation, this is $\|\mu\|_{\m{P}}^{2} = \|\mu\|_{\b{H}(K)}^{2}$, leading to $\m{P}^{\RKHS} = \m{F}^{\RKHS} = \m{K}$. For second-order estimation, the penalty becomes the squared tensor-product norm $\|\Gamma\|_{\m{P}_{\otimes}}^{2}= \|\Gamma\|_{\b{H}(K) \otimes \b{H}(K)}^{2}$, resulting in $\m{P}_{\otimes}^{\RKHS} = \m{F}_{\otimes}^{\RKHS} = \m{O}^{\flat} (\m{K} \odot_{\flat} \m{K}) \m{O}^{\flat}$.
\end{itemize}

With a penalty parameter $\eta \ge 0$, where $\eta = 0$ corresponds to the unregularized problem, the regularized least-squares problem becomes:
\begin{equation*}
    \min_{\m{a} \in \c{C}} \vertii{\m{F} \m{a} - \m{y}}_{\c{D}}^{2} + \eta \innpr{\m{a}}{\m{P} \m{a}}_{\c{C}},
\end{equation*}
which yields the normal equations:
\begin{equation}\label{eq:normal:gen:intro}
    \underbrace{(\m{F}^{*} \m{F} + \eta \m{P})}_{=:\m{S}(\eta)} \hat{\m{a}}(\eta) = \m{F}^{*} \m{y} \, \in \c{C}.
\end{equation}
We call $\m{S}(\eta): \c{C} \to \c{C}$ the regularized Gram matrix, and $\m{F}^{\dagger}(\eta) := [\m{S}(\eta)]^{\dagger} \m{F}^{*} : \c{D} \to \c{C}$ the regularized inverse. For the unregularized case, we omit $\eta=0$ since $\m{F}^{\dagger} = (\m{F}^{*} \m{F})^{\dagger} \m{F}^{*} =  \m{S}^{\dagger} \m{F}^{*}$.

\subsection{Numerical Challenges}
The primary bottleneck in the pool-then-smooth framework is the computational cost of the second-order problem, which far exceeds that of the first-order problem.
The second-order sample size $N_{\otimes} = |\m{r}^{2}|$ scales linearly with $n$ and quadratically with $r_{i}$, leading to acute numerical challenges when $r_{i}$ is large -- a common occurrence in complex scenarios such as spatio-temporal data or higher-dimensional settings, where capturing second-order dependencies requires large $r_{i}$ \cite{gneiting2006geostatistical, sarkar2022covnet, landman2023krylov}. Consequently, when $N_{\otimes}$ becomes prohibitively large, various strategies have been proposed to alleviate computational costs by enforcing sparsity -- trading off estimation fidelity for tractability:
\begin{itemize}[leftmargin=*]
\item In dictionary methods, the second-order normal equation for \eqref{eq:normal:gen:intro} becomes:
\begin{equation}\label{eq:normal:dict:intro}
    \underbrace{[(\m{F}_{\otimes}^{\DICT})^{\top} \m{F}_{\otimes}^{\DICT} + \eta (\m{D}_{\otimes}^{\DICT})^{\top} \m{D}_{\otimes}^{\DICT}]}_{=\m{S}_{\otimes}^{\DICT}(\eta) \in \b{R}^{p^{2} \times p^{2}}} \hat{\m{a}}_{\otimes}^{\DICT}(\eta) = (\m{F}_{\otimes}^{\DICT})^{\top} \m{y}_{\otimes} \in \b{R}^{p^{2}}.
\end{equation}
To \emph{explicitly} store the regularized Gram operator $\m{S}_{\otimes}^{\DICT}(\eta)$ or the regularized inverse $[\m{F}_{\otimes}^{\DICT}]^{\dagger}(\eta)$ -- referred to as sandwich smoothing -- the number of basis functions $p$ is chosen to be small in advance \cite{xiao2013fast, xiao2016fast, xiao2018fast, li2020fast, cederbaum2018fast}. In such cases, $\m{F}_{\otimes}^{\DICT} \in \b{R}^{|\m{r}^{2}| \times p^{2}}$ becomes a tall-and-skinny sparse matrix. This approach implicitly assumes low-rank separability in the covariance function, often justified by a truncated Mercer decomposition \cite{masak2022inference, paul2011principal}.

\item RKHS-based methods take a different route due to their data-adaptive nature:
$\c{C}_{\otimes}^{\RKHS} = \c{D}_{\otimes} = \b{R}^{|\m{r}^{2}|}$. However, as the forward and penalty matrices coincide, the second-order normal equation reduces to
\begin{equation}\label{eq:normal:rkhs:intro}
    (\m{F}_{\otimes}^{\RKHS} + \eta \m{I}) \hat{\m{a}}_{\otimes}^{\RKHS}(\eta) = \m{O}^{\flat} \m{y}_{\otimes} \in \b{R}^{|\m{r}^{2}|}.
\end{equation}
As direct low-rank sparsity with small $p$ is infeasible, \cite{caponera2022functional} instead enforces sparsity directly in the second-order forward matrix $\m{F}_{\otimes}^{\RKHS} = \m{O}^{\flat} (\m{K} \odot_{\flat} \m{K}) \m{O}^{\flat} \in \b{R}^{|\m{r}^{2}| \times |\m{r}^{2}|}$ via aggressive component truncation of the kernel Gram matrix $\m{K} \in \b{R}^{|\m{r}| \times |\m{r}|}$. Otherwise, \emph{explicitly} storing a $|\m{r}^{2}| \times |\m{r}^{2}|$ matrix during preprocessing would require approximately $300$ [GB] of memory when $n=20$ and $r_{i} \equiv 100$. Theoretically, such truncation corresponds to kernel compression via Schur complements, which defines a reproducing kernel for a subspace of the original RKHS $\b{H}(K)$ \cite{paulsen2016introduction}. Low-rank approximation approaches, such as the Nystr\"{o}m method \cite{williams2000using} or incomplete matrix factorizations \cite{bach2002kernel, mahoney2009cur}, can also be applied.
\end{itemize}
While these sparsity-inducing strategies are often effective in practice, they are primarily motivated by computational necessity rather than structural belief. In many applications, the assumption of low-rank separability often fails \cite{aston2017tests, constantinou2017testing, gneiting2006geostatistical}, and aggressive component truncation -- effectively reducing the kernel to a highly localized version -- contradicts the core virtue of smoothing. Although numerical considerations may make dictionary methods appear always appealing -- particularly when $p = o(|\m{r}|)$ -- they are not suitable in settings where the underlying random functions are not directly observed, as in functional inverse problems \cite{yun2025computerized, yun2025low}.

We conclude this section by highlighting the key limitations of existing covariance smoothers. 
First, as shown in \cref{table:LSE:2nd}, evaluating $\m{F}_{\otimes}$ explicitly can be computationally expensive. Moreover, due to the partitioned nature of the Khatri–Rao product, standard factorizations of the first-order forward matrix $\m{F}$ -- such as eigen, Cholesky, or LDL\textsuperscript{T} decompositions -- do not carry over to $\m{F}_{\otimes}$ \cite{liu2008hadamard}. While incomplete block factorizations of $\m{F}_{\otimes}$ can be applied, they typically rely on iterative methods anyway. More importantly, such preprocessing disrupts the structural relationship between $\m{F}_{\otimes}$ and $\m{F}$, leading to inefficiencies.

Second, any efficient Kronecker product operations are implemented at the matrix level, hence flattening the problem to vector form introduces unnecessary obfuscation. This is especially problematic when evaluating $\m{F}_{\otimes}$ explicitly, as the involvement of the elimination matrix $\m{O}_{i}^{\flat}$ depends heavily on the specific observation pattern and indexing scheme for vectorization. While observational schemes $\c{O}_{i}$ may vary in different setups, it is usually straightforward to adjust for this variation, and thus the action of $\s{O}_{i}$ on matrix inputs is simple to adapt. However, its explicit matrix representation depends intricately on the specific reordering of the double array $(j_1, j_2) \notin \c{O}_{i}$, making general implementation more cumbersome.
Additionally, while exploiting symmetry (e.g., through half-vectorization) can reduce computational cost of the linear system by a constant factor, the most substantial performance improvements -- by an order of magnitude -- stem from avoiding explicit evaluation altogether and instead leveraging efficient block tensor-matrix operations, see \cref{sec:cplxty}.

In the following section, we demonstrate how the forward actions  can be efficiently implemented directly at the matrix level, which is sufficient for using Krylov subspace methods. 
Moreover, once the linear system is solved, the estimated surface of $\hat{\Gamma}$ or $\hat{\Sigma}$ is typically evaluated over a square grid, and FPCA is conducted \cite{wang2024adaptive}. These post-processing steps represent the ultimate goals of the estimation procedure. Crucially, the relevant information for these tasks is encoded in the coefficient matrix itself -- not in its vectorized form -- further reinforcing the advantages of a block tensor-matrix formulation.

\section{Block Tensor-Matrix Formulation}\label{sec:cov:est}

\subsection{Block Outer Product}\label{ssec:block:outer}
Given matrices $\m{A} \in \b{R}^{a_{1} \times a_{2}}$ and $\m{B} \in \b{R}^{b_{1} \times b_{2}}$, their outer product $\m{A} \otimes \m{B}$ defines the linear operator:
\begin{equation*}
    (\m{A} \otimes \m{B}) : \b{R}^{b_{2} \times a_{2}} \to \b{R}^{b_{1} \times a_{1}}, \quad \m{C} \mapsto \m{B} \m{C} \m{A}^{\top},
\end{equation*}
which lifts the Kronecker product to the \emph{matrix-level}. When $a_{2} = b_{2} = 1$, this bilinear form entails $\m{a} \otimes \m{b} = \m{b} \m{a}^{\top} \in \b{R}^{b_{1} \times a_{1}}$ for vectors $\m{a} \in \b{R}^{a_{1}}$ and $\m{b} \in \b{R}^{b_{1}}$. The following identities are trivial:
\begin{equation*}
    (\m{A} \otimes \m{B})^{*} = \m{A}^{\top} \otimes \m{B}^{\top}, \quad (\m{A} \otimes \m{B})^{\dagger} = \m{A}^{\dagger} \otimes \m{B}^{\dagger}, \quad (\m{A}_{1} \otimes \m{B}_{1})(\m{A}_{2} \otimes \m{B}_{2}) = (\m{A}_{1} \m{A}_{2} \otimes \m{B}_{1} \m{B}_{2}).
\end{equation*}

Recall $\m{r} = [r_{1}, \dots, r_{n}] \in \b{N}^{n}$ to denote the number of observed locations per sample path. To highlight the block structure of the data, we define the following direct sums:
\begin{equation*}
    \b{R}^{\m{r}} := \bigoplus_{i=1}^{n} \b{R}^{r_{i}}, \quad \odiag(\b{R}^{\m{r}} \otimes \b{R}^{\m{r}}) := \bigoplus_{i=1}^{n} \b{R}^{r_{i} \times r_{i}}.
\end{equation*}
As each space $\b{R}^{r_{i}}$ and $\b{R}^{r_{i} \times r_{i}}$ 
carries a natural Euclidean and Frobenius inner product, their direct sums inherit the natural inner products:
\begin{equation}\label{eq:innpr:direct}
    \innpr{\m{c}}{\m{d}}_{\b{R}^{\m{r}}} = \sum_{i=1}^{n} \m{c}_{i}^{\top} \m{d}_{i}, \quad \innpr{\m{C}}{\m{D}}_{\odiag(\b{R}^{\m{r}} \otimes \b{R}^{\m{r}})} = \sum_{i=1}^{n} \tr (\m{C}_{i}^{\top} \m{D}_{i}).
\end{equation}
Under canonical inner-product isomorphisms, the vectorization of $\odiag(\b{R}^{\m{r}} \otimes \b{R}^{\m{r}})$ becomes:
\begin{equation*}
    \m{C} = \odiag[\m{C}_{i}] =
    \left( \begin{array}{c|c|c|c}
    \m{C}_{1} & \m{0} & \dots & \m{0} \\
    \hline
    \m{0} & \m{C}_{2} & \dots & \m{0} \\
    \hline
    \vdots & \vdots & \ddots & \vdots \\
    \hline
    \m{0} & \m{0} & \dots & \m{C}_{n}
    \end{array} \right) \in \odiag(\b{R}^{\m{r}} \otimes \b{R}^{\m{r}}) 
    \quad \cong \quad
    \begin{pmatrix}
    \ovec(\m{C}_{1}) \\
    \ovec(\m{C}_{2}) \\
    \vdots \\
    \ovec(\m{C}_{n})
    \end{pmatrix} \in \b{R}^{|\m{r}^{2}|}.
\end{equation*}
The data layout of the direct sum can be thought as a (1D) vector of square (2D) matrices of varying sizes. Due to this heterogeneity in dimensions (1D vs. 2D), existing approaches flatten these matrix blocks into (1D) vectors to fit conventional matrix–vector formulations. In contrast, we show that considering this structure as block-diagonal (2D) matrix delivers both conceptual and practical advantages.
To this end, we define the evaluation matrix in dictionary-based methods as a block matrix:
\begin{equation*}
    \m{\Phi} = 
    \left( \begin{array}{c}
    \m{\Phi}_{1} \\
    \hline
    \vdots \\
    \hline
    \m{\Phi}_{n}
    \end{array} \right) \in \b{R}^{\m{r} \times p}: \quad
    \m{c} \in \b{R}^{p} \mapsto 
    \left( \begin{array}{c}
    \m{\Phi}_{1} \m{c} \\
    \hline
    \vdots \\
    \hline
    \m{\Phi}_{n} \m{c}
    \end{array} \right) \in \b{R}^{\m{r}}.
\end{equation*}
Using the inner product in \eqref{eq:innpr:direct}, the adjoint of $\m{\Phi}$ is given by:
\begin{equation*}
    \m{\Phi}^{*} =
    \left( \begin{array}{c|c|c}
    \m{\Phi}_{1}^{\top} &
    \cdots &
    \m{\Phi}_{n}^{\top}
    \end{array} \right) \in \b{R}^{p \times \m{r}}: \quad 
    \left( \begin{array}{c}
    \m{c}_{1} \\
    \hline
    \vdots \\
    \hline
    \m{c}_{n}
    \end{array} \right) \in \b{R}^{\m{r}} \mapsto \sum_{i=1}^{n} \m{\Phi}_{i}^{\top} \m{c}_{i} \in \b{R}^{p}.
\end{equation*}
Similarly, the kernel Gram matrix in RKHS-based methods is expressed with block structure as:
\begin{align*}
    \m{K} = 
    \left( \begin{array}{c}
    \m{K}_{1\cdot} \\
    \hline
    \vdots \\
    \hline
    \m{K}_{n\cdot}
    \end{array} \right) =
    \left( \begin{array}{c|c|c}
    \m{K}_{11} & \cdots & \m{K}_{1n} \\
    \hline
    \vdots & \ddots & \vdots \\
    \hline
    \m{K}_{n1} & \cdots & \m{K}_{nn}
    \end{array} \right) \in \b{R}^{\m{r} \times \m{r}}: \quad
    \m{c} = 
    \left( \begin{array}{c}
    \m{c}_{1} \\
    \hline
    \vdots \\
    \hline
    \m{c}_{n}
    \end{array} \right) \in \b{R}^{\m{r}} \mapsto 
    \left( \begin{array}{c}
    \m{K}_{1\cdot} \m{c} \\
    \hline
    \vdots \\
    \hline
    \m{K}_{n\cdot} \m{c}
    \end{array} \right) \in \b{R}^{\m{r}}.
\end{align*}

We now define the Khatri–Rao product at the matrix level. First, the row-wise outer product is given by
\begin{equation*}
    \m{\Phi} \odot \m{\Phi} 
    : \quad \m{C} \in \b{R}^{p \times p} \mapsto \odiag[\m{\Phi}_{i} \m{C} \m{\Phi}_{i}^{\top}] \in \odiag(\b{R}^{\m{r}} \otimes \b{R}^{\m{r}}). 
\end{equation*}
In the special case $p = 1$, this corresponds to the observation vector $\m{y}_{\otimes}$ in \eqref{eq:data:intro} upon vectorization:
\begin{equation*}
    \m{y} = 
    \left( \begin{array}{c}
    \m{y}_{1} \\
    \hline
    \vdots \\
    \hline
    \m{y}_{n}
    \end{array} \right) \in \b{R}^{\m{r}} \quad \Longrightarrow \quad
    \m{y} \odot \m{y} = \odiag[\m{y}_{i} \m{y}_{i}^{\top}] \in \odiag(\b{R}^{\m{r}} \otimes \b{R}^{\m{r}}).
\end{equation*}
The adjoint yields the column-wise outer product:
\begin{equation*}
    (\m{\Phi} \odot \m{\Phi})^{*} = \m{\Phi}^{*} \odot \m{\Phi}^{*} 
    : \quad  \odiag[\m{C}_{i}] \in \odiag(\b{R}^{\m{r}} \otimes \b{R}^{\m{r}}) \mapsto \sum_{i=1}^{n} \m{\Phi}_{i}^{\top} \m{C}_{i} \m{\Phi}_{i} \in \b{R}^{p \times p}. 
\end{equation*}
Finally, the block-wise outer product of the kernel Gram matrix $\m{K} \in \b{R}^{\m{r} \times \m{r}}$ is defined by:
\begin{equation*}
    \m{K} \odot \m{K} : \quad \m{C} = \odiag[\m{C}_{i}] \in \odiag(\b{R}^{\m{r}} \otimes \b{R}^{\m{r}}) \mapsto \odiag[\m{K}_{i \cdot} \m{C} \m{K}_{\cdot i}] \in \odiag(\b{R}^{\m{r}} \otimes \b{R}^{\m{r}}),
\end{equation*}
where $\m{K}_{i \cdot} \in \b{R}^{r_{i} \times \m{r}}$ and $\m{K}_{\cdot i} := \m{K}_{i \cdot}^{*} \in \b{R}^{\m{r} \times r_{i}}$ for $1 \le i \le n$. Since $\m{K} \succeq \m{0}$, it is straightforward that $(\m{K}_{i\cdot} \otimes \m{K}_{i\cdot}) \succeq \m{0}$ for each $1 \le i \le n$, leading to $\m{K} \odot \m{K} \succeq \m{0}$.
As the row- and column-wise outer products for dictionary methods are even more straightforward to implement, we present the pseudocode of the block-wise outer product for RKHS methods in \cref{alg:lazy:khatri:block}.



\begin{algorithm}[h!]
\caption{Block-wise Outer Product}\label{alg:lazy:khatri:block}
\begin{algorithmic}[1]
\Require $\m{K} \in \b{R}^{\m{r} \times \m{r}}, \m{C} = \odiag[\m{C}_{i}] \in \odiag(\b{R}^{\m{r}} \otimes \b{R}^{\m{r}})$
\Ensure $\m{D}= (\m{K} \odot \m{K}) \m{C} \in \odiag(\b{R}^{\m{r}} \otimes \b{R}^{\m{r}})$
    \For{$i = 1$ to $n$}
        \State $\m{D}_i \gets \m{0}$
        \For{$i' = 1$ to $n$}
            \State $\m{D}_i \gets \m{D}_i + \m{K}_{i i'} \m{C}_{i'} \m{K}_{i i'}^\top$
        \EndFor
    \EndFor    
\end{algorithmic}
\end{algorithm}

The elimination operator acts block-wise as follows, see \cref{alg:diag:elim} for pseudocode: 
\begin{equation*}
    \s{O} = \odiag[\s{O}_{i}] :  \quad \odiag[\m{C}_{i}] \in \odiag(\b{R}^{\m{r}} \otimes \b{R}^{\m{r}}) \mapsto \odiag[\s{O}_{i} \m{C}_{i}] \in \odiag(\b{R}^{\m{r}} \otimes \b{R}^{\m{r}}),
\end{equation*}
which is an orthogonal projection operator.
Note that, if the input is a symmetric (partitioned) matrix, then the output of any of the operators -- $\m{\Phi} \odot \m{\Phi}$, $\m{\Phi}^{*} \odot \m{\Phi}^{*}$, $\m{K} \odot \m{K}$, or $\s{O}$ -- also preserves symmetry. Consequently, one can exploit symmetry to reduce memory usage by storing only the upper (or lower) triangular portion of the block diagonal matrices. While this reduction is theoretically appealing, it does not constitute a major computational bottleneck and not necessarily yields an expected amount of speedup, see \cref{sec:cplxty}.

\begin{algorithm}[h!]
\caption{Diagonal Elimination}\label{alg:diag:elim}
\begin{algorithmic}[1]
\Require $\m{C} = \odiag[\m{C}_{i}] \in \odiag(\b{R}^{\m{r}} \otimes \b{R}^{\m{r}})$
\Ensure $\m{D} = \s{O} \m{C} \in \odiag(\b{R}^{\m{r}} \otimes \b{R}^{\m{r}})$
    \State $\m{D} \gets \m{C}$
    \For{$i = 1$ to $n$}
        \For{$j = 1$ to $r_i$}
            \State $\m{D}_i[j, j] \gets 0$
        \EndFor
    \EndFor    
\end{algorithmic}
\end{algorithm}

\subsection{Ridge Regression}
The Tikhonov loss for the mean function $\mu$ is defined as
\begin{equation}\label{eq:loss:1st}
    \c{L}^{\eta}(\mu) = \sum_{i=1}^{n} \sum_{j=1}^{r_{i}} (y_{ij} - \mu(t_{ij}))^{2} + \eta \|\mu\|_{\m{P}}^{2}.
\end{equation}

When $\mu \in \c{S}$ is represented by a coefficient vector $\m{a}$, we also write $\c{L}^{\eta}(\m{a})$. The minimizers are denoted by $\hat{\mu}(\eta)$ and $\hat{\m{a}}(\eta)$, where we omit the dependency on $\eta$ when $\eta = 0$ (unregularized case). We denote by $\c{G} := \{t_{m} : 1 \le m \le g\} \subset \c{T}$ the grid over which $\hat{\mu}(\eta): \c{T} \to \b{R}$ is evaluated.
Similarly, the Tikhonov loss for the second-moment function $\Gamma$ is given by
\begin{align}\label{eq:loss:2nd}
    \c{L}_{\otimes}^{\eta}(\Gamma)
    = \sum_{i=1}^{n} \sum_{(j_{1}, j_{2}) \in \c{O}_{i}} \left(y_{ij_{1}} y_{ij_{2}} -  \Gamma(t_{ij_{1}}, t_{ij_{2}}) \right)^{2} + \eta \|\Gamma\|_{\s{P}}^{2}, 
\end{align}
If $\Gamma \in \c{S}_{\otimes}$ is represented by a coefficient matrix $\m{A}$, we equivalently write $\c{L}_{\otimes}^{\eta}(\m{A})$. The corresponding minimizers are denoted by $\hat{\Gamma}(\eta)$ and $\hat{\m{A}}(\eta)$, again omitting $\eta$ when zero. The estimator $\hat{\Gamma}(\eta): \c{T} \times \c{T} \to \b{R}$ is evaluated over the square grid $\c{G} \times \c{G}$.

From this point onward, in the context of block tensor-matrix formulations for second-order estimation, we adopt the following script-style notation: the identity operator is denoted by $\s{I}$, the forward operator by $\s{F}$, and the penalty operator by $\s{P}$ where $\vertii{\Gamma}_{\s{P}}^{2} = \vertii{\m{A}}_{\s{P}}^{2} = \innpr{\m{A}}{\s{P}\m{A}}$. In analogy to \eqref{eq:normal:gen:intro}, we denote the regularized Gram operator by $\s{S}(\eta) = \s{F}^{*} \s{F} + \eta \s{P}$, and the regularized inverse by $\s{F}^{\dagger}(\eta) = [\s{S}(\eta)]^{\dagger} \s{F}^{*}$. For the unregularized case, we omit the dependence on $\eta=0$.

\medskip
\noindent \textbf{Dictionary Methods}

Define the frame matrix $\m{E} \in \b{R}^{g \times p}$ with respect to the evaluation grid $\c{G}$ by $\m{E}[m, l]= \phi_{l}(t_{m})$ \cite{belytschko1994element}. Then, the evaluation of
\begin{equation*}
    \mu = \sum_{l=1}^{p} a_{l} \phi_{l}, \quad \m{a} = [a_{1}, \cdots, a_{p}]^{\top} \in \b{R}^{p},
\end{equation*}
is given by $\boldsymbol{\mu} := [\mu(t_{m})]_{1 \le m \le g} = \m{E} \m{a} \in \b{R}^{g}$. Also, the loss functional in \eqref{eq:loss:1st} becomes
\begin{align*}
    \c{L}^{\eta}(\m{a}) = \sum_{i=1}^{n} \vertii{\m{y}_{i} - \m{\Phi}_{i} \m{a}}^{2} + \eta \vertii{\m{a}}_{\m{P}}^{2} = \vertii{\m{y} - \m{\Phi} \m{a}}^{2} + \eta \vertii{\m{a}}_{\m{P}}^{2},
\end{align*}
which leads to the normal equation:
\begin{equation*}
    (\m{\Phi}^{*} \m{\Phi} +\eta \m{P}) \hat{\m{a}}(\eta) = \m{\Phi}^{*} \m{y} \quad \Longleftrightarrow \quad \left( \sum_{i=1}^{n} \m{\Phi}_{i}^{*} \m{\Phi}_{i}  +\eta \m{P} \right) \hat{\m{a}}(\eta) = \sum_{i=1}^{n} \m{\Phi}_{i}^{*} \m{y}_{i}.
\end{equation*}
In the rank-deficient case, the solution of minimum norm is given by $\hat{\m{a}}(\eta) = \m{\Phi}^{\dagger}(\eta) \m{y} \in \b{R}^{p}$, where $\m{\Phi}^{\dagger}(\eta)$ is the regularized inverse defined in \eqref{eq:normal:gen:intro}.

For second-moment estimation, we consider
\begin{equation*}
    \Gamma = \sum_{l_{1}, l_{2} = 1}^{p} a_{l_{1} l_{2}} \phi_{l_{1}} \otimes \phi_{l_{2}}, \quad \m{A} \in \b{R}^{p \times p},
\end{equation*}
where its evaluation over the square grid $\c{G} \times \c{G}$ is given by
\begin{equation*}
    \boldsymbol{\Gamma} := [\Gamma(t_{m_{1}}, t_{m_{2}})]_{1 \le m_{1}, m_{2} \le g} = (\m{E} \otimes \m{E}) \m{A} = \m{E} \m{A} \m{E}^{\top} \in \b{R}^{g \times g}.
\end{equation*}
It readily follows that the evaluation of the covariance $\Sigma$ over the square grid $\c{G} \times \c{G}$ is given by
\begin{equation*}
    \boldsymbol{\Sigma} := [\Sigma(t_{m_{1}}, t_{m_{2}})]_{1 \le m_{1}, m_{2} \le g} = (\m{E} \otimes \m{E}) (\m{A} - \m{a} \otimes \m{a}) = \m{E} (\m{A} - \m{a} \m{a}^{\top}) \m{E}^{\top} \in \b{R}^{g \times g}.
\end{equation*}

\begin{theorem}\label{thm:dict:2nd:mom}
Denote the identity operator by $\s{I} : \odiag(\b{R}^{\m{r}} \otimes \b{R}^{\m{r}}) \to \odiag(\b{R}^{\m{r}} \otimes \b{R}^{\m{r}})$, the forward operator by $\s{F} := \s{O} (\m{\Phi} \odot \m{\Phi}) : \b{R}^{p \times p} \to \odiag(\b{R}^{\m{r}} \otimes \b{R}^{\m{r}})$ and the s.p.d. penalty operator by $\s{P}: \b{R}^{p \times p} \to \b{R}^{p \times p}$. Then, the loss functional in \eqref{eq:loss:2nd} becomes
\begin{align*}
    \c{L}_{\otimes}^{\eta}(\m{A})
    = \vertii{(\m{y} \odot \m{y}) - \s{F} \m{A}}^{2} + \eta \vertii{\m{A}}_{\s{P}}^{2} - \vertii{(\s{I} - \s{O}) (\m{y} \odot \m{y})}^{2}.
\end{align*}
The last term does not depend on $\m{A}$, and the normal equation is given by
\begin{align}\label{eq:dict:2nd:normal}
    (\s{F}^{*} \s{F} + \eta \s{P}) \hat{\m{A}}(\eta) = \s{F}^{*} (\m{y} \odot \m{y}).
\end{align}
\end{theorem}

Again, in the rank-deficient case, the solution of minimum Frobenius norm is given by $\hat{\m{A}}(\eta) = \s{F}^{\dagger}(\eta) (\m{y} \odot \m{y}) \in \b{R}^{p \times p}$, which is symmetric due to the convexity of $\c{L}_{\otimes}^{\eta}$ for any $\eta \ge 0$.

\medskip
\noindent \textbf{RKHS Methods}

In the RKHS-based approach \cite{cai2010nonparametric, cai2011optimal, caponera2022functional, yun2025computerized, unser2021unifying, yun2025low}, we consider $n$ i.i.d. copies $\{X_{i}: i=1, \dots, n \}$ to be a second-order random element in $\b{H}$. The mean $\mu$, second moment tensor $\Gamma$, and the covariance tensor $\Sigma$ of $X_{1}$ are defined by Bochner integral \cite{hsing2015theoretical}:
\begin{align}\label{eq:tens=ftns}
    \mu := \b{E} X_{1} \in \b{H}, \quad
    \Gamma := \b{E} [X_{1} \otimes X_{1}] \in \b{H} \otimes \b{H}, \quad
    \Sigma := \Gamma - \mu \otimes \mu \in \b{H} \otimes \b{H},
\end{align}
which are same when considered as functions in \eqref{eq:def:moments} due to the reproducing property \eqref{eq:tens:repro:prop}.

The representer theorem \cite{kimeldorf1970correspondence, wahba1981spline} asserts that the search space becomes $\c{S} = \spann \{\rk_{ij} : 1 \le i \le n, 1 \le j \le r_{i} \}$. Unlike dictionary-based methods, the representer theorem yields a data-driven basis \textit{a posteriori}, tailored to the observed input locations. In essence, any finite-dimensional inner product space is an RKHS \cite{paulsen2016introduction}, so the RKHS methods can be seen as spline methods, where \emph{all} the observed locations effectively serve as knots. In this regard, we consider the coefficient vector to be a direct sum:
\begin{equation*}
    \mu = \sum_{i=1}^{n} \sum_{j=1}^{r_{i}} a_{ij} \rk_{ij} \in \c{S}, \quad \m{a} = 
    \left( \begin{array}{c}
    \m{a}_{1} \\
    \hline
    \vdots \\
    \hline
    \m{a}_{n}
    \end{array} \right) \in \b{R}^{\m{r}},
\end{equation*}
which leads to the normal equation $(\m{K} + \eta \m{I}) \hat{\m{a}}(\eta) = \m{y}$. Note that the solution is unique when $\eta > 0$ due to the strict convexity of $\c{L}^{\eta}$ in \eqref{eq:loss:1st}, regardless of rank-deficiency of the kernel Gram matrix $\m{K} \succeq \m{0}$. In the case where $\eta = 0$, we can impose the uniqueness by choosing $\hat{\mu}$ of minimum norm, which is given by $\hat{\m{a}} = \m{K}^{\dagger} \m{y}$ \cite{paulsen2016introduction}.

Define the frame block matrix $\m{E} = ( \m{E}_{1} \vert \cdots \vert \m{E}_{n} ) \in \b{R}^{g \times \m{r}}$, where the $i$-th block $\m{E}_{i} \in \b{R}^{g \times r_{i}}$ is given by $\m{E}_{i}[m, j]= K(t_{m}, t_{ij})$. Then, the evaluation of $\mu \in \b{H}$ over $\c{G}$ is given by
\begin{equation*}
    \boldsymbol{\mu} = \m{E} \m{a} = \sum_{i=1}^{n} \m{E}_{i} \m{a}_{i} \in \b{R}^{g}.
\end{equation*}

Similarly, the search space for second-moment estimation guided by the representer theorem becomes $\spann \{\rk_{ij_{1}} \otimes \rk_{ij_{2}} : 1 \le i \le n, (j_{1}, j_{2}) \in \c{O}_{i} \}$. However, it is rather convenient to impose a restriction on the coefficients:
\begin{equation*}
    \Gamma = \sum_{i=1}^{n} \sum_{j_{1}, j_{2} =1}^{r_{i}} a_{ij_{1}j_{2}} \rk_{ij_{1}} \otimes \rk_{ij_{2}}, \quad \m{A} = \odiag[\m{A}_{i}] \in \c{R}(\s{O}^{\oplus}) \subset \odiag(\b{R}^{\m{r}} \otimes \b{R}^{\m{r}}).
\end{equation*}
Consequently, the evaluation of $\Gamma$ and $\Sigma \in \b{H} \otimes \b{H}$ over the square grid $\c{G} \times \c{G}$ are given by the column-wise outer product:
\begin{align*}
    &\boldsymbol{\Gamma} = \m{E} \m{A} \m{E}^{*}  = \sum_{i=1}^{n} \m{E}_{i} \m{A}_{i} \m{E}_{i}^{\top} \in \b{R}^{g \times g}, \quad
    \boldsymbol{\Sigma} = \m{E} (\odiag[\m{A}_{i}] - \m{a} \m{a}^{*}) \m{E}^{*} \in \b{R}^{g \times g}.  
\end{align*}

\begin{theorem}\label{thm:rkhs:2nd:mom}
Denote the forward operator on $\odiag(\b{R}^{\m{r}} \otimes \b{R}^{\m{r}})$ by $\s{F} := \s{O} (\m{K} \odot \m{K}) \s{O}$. Then, the loss functional in \eqref{eq:loss:2nd} becomes
\begin{align*}
    \c{L}_{\otimes}^{\eta}(\m{A})
    = \vertii{(\m{y} \odot \m{y}) - \s{F} \m{A}}^{2} + \eta \vertii{\m{A}}_{\s{P}}^{2} - \vertii{(\s{I} - \s{O}) (\m{y} \odot \m{y})}^{2}.
\end{align*}
The last term does not depend on $\m{A}$, and the normal equation is given by
\begin{align}\label{eq:rkhs:2nd:normal}
    \s{F}^{*} (\s{F} + \eta \s{I}) \hat{\m{A}}(\eta) = \s{F}^{*} (\m{y} \odot \m{y}).
\end{align}
\end{theorem}

It is important to observe that
\begin{equation}\label{eq:rk:deficient}
    \mathrm{rank} (\s{O}) = \tr (\s{O}) = \sum_{i=1}^{n} r_{i} (r_{i}-1) < |\m{r}^{2}| = \dim [\odiag(\b{R}^{\m{r}} \otimes \b{R}^{\m{r}})],
\end{equation}
which implies that the forward operator $\s{F} := \s{O} (\m{K} \odot \m{K}) \s{O} \succeq \m{0}$ is always rank-deficient. Consequently, the solution to this ill-posed problem \eqref{eq:rkhs:2nd:normal} is not unique in the unregularized case. Among the infinitely many solutions, one may select the minimum-norm solution (both in $\hat{\Gamma}$ and $\hat{\m{A}}$), given by $\hat{\m{A}} = \s{F}^{\dagger} (\m{y} \odot \m{y})$ \cite{paulsen2016introduction}. However, solving this equation directly or iteratively using the Conjugate Gradient (CG) method can lead to numerical instability in the presence of rank deficiency \cite{hanke2017conjugate, hansen1998rank}. In such cases, a more suitable Krylov subspace method, such as MINRES, should be employed \cite{paige1975solution}.
In contrast, when $\eta > 0$, the solution to \eqref{eq:rkhs:2nd:normal} is still not unique at the level of the coefficient matrix $\hat{\m{A}}(\eta)$. However, all such solutions correspond to the same function $\hat{\Gamma}(\eta)$, owing to the strict convexity of the Tikhonov loss $\c{L}_{\otimes}^{\eta}$ defined in \eqref{eq:loss:2nd} \cite{paulsen2016introduction}. Therefore, we may equivalently solve the regularized system
\begin{align}\label{eq:rkhs:2nd:normal:simp}
    (\s{F} + \eta \s{I}) \hat{\m{A}}(\eta) = \s{O} (\m{y} \odot \m{y}), \quad \eta \ge 0.
\end{align}

\subsection{Functional Principal Components}\label{ssec:fpca}
Given an estimator of covariance function $\Sigma$, one can associate a corresponding covariance operator on a Hilbert space, and FPCA involves extracting its top eigenvalues and eigenfunctions. These spectral components are encoded in the coefficient matrix $\m{A}$ with respect to the inner-product structure determined by the basis of the search space. However, since this basis is typically non-orthonormal, the problem naturally leads to a \emph{generalized} eigenvalue problem.

\medskip
\noindent \textbf{Dictionary Methods}

The canonical approach associates the covariance function $\Sigma$ with the integral operator
\begin{equation*}
    \s{T}_{\Sigma}: \c{L}_2(\c{T}) \to \c{L}_2(\c{T}), \quad  \s{T}_{\Sigma} f (s) = \int_{\c{T}} \Sigma(s, t) f(t) dt,
\end{equation*}
which is a trace-class s.p.d. operator, and its spectral decomposition yields the celebrated Mercer expansion \cite{hsing2015theoretical, paulsen2016introduction}. When restricting to the search space $\c{S}^{\DICT} = \spann \{\phi_{l}: \c{T} \to \b{R}, 1 \le l \le p \}$, the spectral information is encoded in the generalized eigenvalues and eigenvectors of the centered coefficient matrix $\tilde{\m{A}} := \m{A} - \m{a} \m{a}^{\top} \in \b{R}^{p \times p}$ \cite{xiao2018fast}.

\begin{proposition}\label{thm:fpca:dict}
Let $\m{G} = [g_{l_{1} l_{2}}]_{1 \le l_{1}, l_{2} \le p} \in \b{R}^{p \times p}$ where $g_{l_{1} l_{2}} = \innpr{\phi_{l_{1}}}{\phi_{l_{2}}}_{\c{L}_2(\c{T})}$, and consider the covariance function of the form
\begin{align*}
    \s{T}_{\Sigma} = \sum_{l_{1}, l_{2}=1}^{p} \tilde{a}_{l_{1} l_{2}} \phi_{l_{1}} \otimes_{\c{L}_{2}} \phi_{l_{2}} \quad \Rightarrow \quad \Sigma(s, t) = \sum_{l_{1}, l_{2}=1}^{p} \tilde{a}_{l_{1} l_{2}} \phi_{l_{1}}(s) \phi_{l_{2}}(t), \quad \tilde{\m{A}} \in \b{R}^{p \times p}.
\end{align*}
Let $\lambda_{k}$ and $f_{k} = \sum_{l=1}^{p} v_{lk} \phi_{l}$ denote the $k$-th eigenvalue and eigenfunction of $\Sigma$, i.e. $\Sigma = \sum_{k=1}^{p} \lambda_{k} f_{k} \otimes f_{k}$. If we define
\begin{equation*}
    \m{\Lambda} = \odiag[\lambda_{k}]_{1 \le k \le p} \in \b{R}^{p \times p}, \quad \m{V} = [v_{lk}]_{1 \le l, k \le p} = [ \m{v}_{1} \vert \cdots \vert \m{v}_{p} ] \in \b{R}^{p \times p},
\end{equation*}
then the diagonal matrix $\m{\Lambda}$ and $\m{G}$-orthogonal matrix $\m{V}$ satisfy 
\begin{equation}\label{eq:fpca:rkhs}
    \tilde{\m{A}} \m{G} \m{V} = \m{V} \m{\Lambda}, \quad \m{V}^{\top} \m{G} \m{V} = \m{I}.
\end{equation}
Also, when $\m{G}$ is rank-deficient, the FPCA does not depend on the choice of $\m{V}$.
\end{proposition}

\medskip
\noindent \textbf{RKHS Methods}

In the RKHS setting, while a Mercer expansion in \cref{thm:fpca:dict} is still valid, computing $\c{L}_2(\c{T})$-inner products between feature maps  is typically intractable unless the Sobolev-type kernel is constructed through a concrete differential operator \cite{caponera2022functional, cai2010nonparametric}. Instead, we reinterpret $\Sigma$ as a tensor in $\b{H} \otimes \b{H}$ (as noted in \eqref{eq:tens=ftns}), where the FPCA is then performed via the centered coefficient matrix $\tilde{\m{A}} := \m{A} - \m{a} \m{a}^{*} \in \b{R}^{\m{r} \times \m{r}}$. Note that $\m{a} \m{a}^{*}$ is generically full-block unless identically zero, and hence so is $\tilde{\m{A}}$. Furthermore, since $\m{A} \in \c{R}(\s{O})$ is trace-free, it cannot be s.p.d. unless zero. This degeneracy, not explicitly known in the literature, is closely related to the rank-deficiency discussed in \eqref{eq:rk:deficient}, which highlights the necessity of performing FPCA to truncate negative spectrum tails. Finally, the following PCA is a decomposition in the RKHS, often called the kernel PCA, not a Mercer expansion in $\c{L}_{2}(\c{T})$.

\begin{proposition}\label{thm:fpca:rkhs}
Let $\m{K} \in \b{R}^{\m{r} \times \m{r}}$ be the mean Gram matrix, and consider the covariance tensor of the form
\begin{align*}
    \Sigma = \sum_{i_{1}, i_{2} = 1}^{n} \sum_{j_{1}=1}^{r_{i_{1}}} \sum_{j_{2}=1}^{r_{i_{2}}} \tilde{a}_{i_{1} j_{1}, i_{2} j_{2}} \rk_{i_{1} j_{1}} \otimes_{\b{H}} \rk_{i_{2} j_{2}} \quad \Rightarrow \quad \Sigma(s, t) = \sum_{i_{1}, i_{2} = 1}^{n} \sum_{j_{1}=1}^{r_{i_{1}}} \sum_{j_{2}=1}^{r_{i_{2}}} \tilde{a}_{i_{1} j_{1}, i_{2} j_{2}} \rk_{i_{1} j_{1}}(s) \rk_{i_{2} j_{2}}(t),
\end{align*}
with $\tilde{\m{A}} \in \b{R}^{\m{r} \times \m{r}}$ symmetric. Let $\lambda_k$ and $f_k = \sum_{i=1}^{n} \sum_{j=1}^{r_i} v_{ij, k} \rk_{ij}$ denote the $k$-th eigenvalue and eigenfunction of $\Sigma$ in $\b{H}$, where $1 \le k \le |\m{r}|$. Then defining the diagonal matrix $\m{\Lambda} \in \b{R}^{\m{r} \times \m{r}}$ and $\m{K}$-orthogonal matrix $\m{V} \in \b{R}^{\m{r} \times \m{r}}$ as in \cref{thm:fpca:dict}, we have
\begin{equation*}
    \tilde{\m{A}} \m{K} \m{V} = \m{V} \m{\Lambda}, \quad \m{V}^{*} \m{K} \m{V} = \m{I}. 
\end{equation*}
In the case where $\m{K}$ is rank-deficient, the FPCA does not depend on the choice of $\m{V}$.
\end{proposition}

Both \cref{thm:fpca:dict,thm:fpca:rkhs} lead to the same formulation, so we discuss \eqref{eq:fpca:rkhs} for notational convenience. Let $\m{L}$ be any square matrix satisfying $\m{G} = \m{L} \m{L}^{\top}$. A common two-step factorization-based method \cite{xiao2016fast, caponera2022functional, ramsay2002applied} solves \eqref{eq:fpca:rkhs} by first computing the standard eigendecomposition $\m{L}^{\top} \tilde{\m{A}} \m{L} = \m{U} \m{\Lambda} \m{U}^{\top}$, and then recover the generalized eigenvectors $\m{V}$ via $\m{U} = \m{L}^{\top} \m{V}$. While the final solution is independent of the specific choice (e.g., Cholesky) of $\m{L}$ since all such $\m{L}^{\top} \tilde{\m{A}} \m{L}$ are unitarily equivalent \cite{conway1994course}, a prevalent choice in the literature is $\m{L} = \m{G}^{1/2}$. However, computing the square root requires an additional eigendecomposition of $\m{G}$, which can be computationally intensive and numerically unstable when the spectrum has small gaps \cite{nocedal1999numerical, golub2013matrix}. Moreover, while this method is tractable for moderate matrix sizes, e.g., $\tilde{\m{A}} \in \b{R}^{p \times p}$ in dictionary-based approaches, it becomes impractical in RKHS-based methods, where the kernel matrix $\m{K}$ is large and often ill-conditioned.

From a theoretical perspective, this two-step procedure amounts to orthogonalizing the basis of the search space, performing PCA in that new \emph{artificial} function space, and then mapping the results back to the original basis. However, this detour may obscure the conceptual elegance of the Hilbert space formulation. To maintain abstraction and computational efficiency, we next describe how to compute an incomplete generalized eigendecomposition to approximate the leading principal components iteratively.

\medskip
\noindent \textbf{Lanczos Tridiagonalization}

The Lanczos algorithm \cite{lanczos1950iteration, paige1972computational} is a cornerstone of Krylov subspace methods for symmetric matrices $\tilde{\m{A}}$, where the original least-squares problem is projected over lower-dimensional subspaces involving a tridiagonal matrix $\m{T}_{k}$. We showcase how this algorithm can be tailored into our problem \eqref{eq:fpca:rkhs}, where $\m{G} = \m{I}$ corresponds to the usual algorithm. Once again, especially for the RKHS methods, we do not need to explicitly evaluate the full block matrix $\tilde{\m{A}} = \m{A} - \m{a} \m{a}^{*} \in \b{R}^{\m{r} \times \m{r}}$, but rather implement its action on $\b{R}^{\m{r}}$.

Starting with an initial vector $\m{b}_{0} \in \b{R}^{p}$, we construct the Krylov subspace for the matrix $\tilde{\m{A}} \m{G}$:
\begin{equation*}
    \c{K}_{k}(\tilde{\m{A}} \m{G}, \m{b}_{0}) = \spann \{ \m{b}_{0}, (\tilde{\m{A}} \m{G}) \m{b}_{0}, \cdots, (\tilde{\m{A}} \m{G})^{k-1} \m{b}_{0} \}.
\end{equation*}
Note that $\tilde{\m{A}} \m{G}$ is $\m{G}$-symmetric since $\langle \m{c}, \tilde{\m{A}} \m{G} \m{d} \rangle_{\m{G}} = \langle \tilde{\m{A}} \m{G} \m{c}, \m{d} \rangle_{\m{G}}$. The $\m{G}$-Lanczos procedure generates a basis for this subspace using a set of vectors that are orthogonal with respect to the $\m{G}$-inner product. After $k$ iterations, as outlined in \cref{alg:lanczos}, this process yields a $\m{G}$-orthonormal matrix $\m{U}_{k} \in \b{R}^{p \times k}$ that spans $\c{K}_{k}(\tilde{\m{A}} \m{G}, \m{b}_{0})$, satisfying the three-term recurrence \cite{nocedal1999numerical}:
\begin{equation*}
    \tilde{\m{A}} \m{G} \m{U}_{k} = \m{U}_{k} \m{T}_{k} + \beta_{k} \m{u}_{k+1} \m{e}_{k}^{\top}, \quad \m{T}_{k} =
    \begin{pmatrix}
    \alpha_{1} & \beta_{1} & 0 & \cdots & 0 \\
    \beta_{1} & \alpha_{2} & \beta_{2} & \ddots & 0 \\
    0 &\beta_{2} & \alpha_{3} & \ddots & \vdots \\
    \vdots & \ddots & \ddots & \ddots & \vdots \\
    0 & \cdots & 0 & \beta_{k-1} & \alpha_{k}
    \end{pmatrix} \in \b{R}^{k \times k}, \quad \m{e}_{k} = \begin{pmatrix}
        0 \\ 0 \\ \vdots \\ 0 \\ 1
    \end{pmatrix} \in \b{R}^{k},
\end{equation*}
where the second term is the residual. Left-multiplying by $\m{U}_{k}^{\top} \m{G}$, we obtain the projected system:
\begin{equation*}
    \m{U}_{k}^{\top} [\m{G} \tilde{\m{A}} \m{G}] \m{U}_{k} = [\m{U}_{k}^{\top} \m{G} \m{U}_{k}] \m{T}_{k} + \beta_{k} [\m{U}_{k}^{\top} \m{G} \m{u}_{k+1}] \m{e}_{k}^{\top} = \m{T}_{k}.
\end{equation*}
The eigenpairs $(\m{\Lambda}_k, \m{S}_k)$ of this small projected system $\m{T}_{k} = \m{S}_{k} \m{\Lambda}_{k} \m{S}_{k}^{\top}$ are known as the Ritz values and Ritz vectors, respectively. We can then form approximations to the generalized eigenvectors of the original problem by mapping the Ritz vectors back to the high-dimensional space by letting $\m{V}_{k} = \m{U}_{k} \m{S}_{k} \in \b{R}^{p \times k}$. These Ritz vectors $\m{V}_{k} = [\m{v}_{1} \vert \m{v}_{2} \vert \cdots \vert \m{v}_{k}]$ are $\m{G}$-orthonormal by construction:
\begin{equation*}
    \m{V}_{k}^{\top} \m{G} \m{V}_{k} = \m{S}_{k}^{\top} (\m{U}_{k}^{\top} \m{G} \m{U}_{k}) \m{S}_{k} = \m{S}_{k}^{\top} \m{I}_{k} \m{S}_{k} = \m{I}_{k}.
\end{equation*}
In exact arithmetic, at most $\mathrm{rank}( \tilde{\m{A}} \m{G})$ iterations yield the exact solutions. The quality of the approximation, which is assessed by the $\m{G}$-norm of the $j$-th residual vector
\begin{equation*}
    \vertii{\m{r}_{j}}_{\m{G}} = \| \tilde{\m{A}} \m{G} \m{v}_{j} - \lambda_{k} \m{v}_{j} \|_{\m{G}} = \vertii{\beta_{k} \m{u}_{k+1} (\m{e}_{k}^{\top} \m{s}_{j})}_{\m{G}} = \beta_{k} |s_{kj}|,
\end{equation*}
converges rapidly to zero as $k$ increases for leading eigenvalues, hence the full iteration is not only unnecessary but may also degrade numerical stability \cite{jia2020low}. Consequently, if we aim to find the Ritz pairs $(\m{\Lambda}_k, \m{V}_k)$ corresponding to the largest eigenvalues, and we only need to compute the first few eigenpairs of $\m{T}_{k}$ to obtain accurate approximations after early termination \cite{chung2024computational}. 

\begin{algorithm}[h!]
\caption{Incomplete $\m{G}$-eigendecomposition}\label{alg:lanczos}
\begin{algorithmic}[1]
\Require $\tilde{\m{A}}, \m{G} \in \b{R}^{p \times p}$
\Ensure $([\lambda_{1}, \cdots, \lambda_{l}], [\m{v}_{1} \vert \cdots \vert \m{v}_{l}])$\Comment{Find $l$ Ritz pairs after $k$ iterations}
\Procedure{$\m{G}$-eigen}{$l, \tilde{\m{A}}, \m{G}; k, \m{b}_{0}, \text{tol}>0$}
    \State $\beta_{0} \gets 0, \m{u}_{0} \gets \m{0}$
    \State $\m{u}_{1} \gets \m{b}_{0} / \vertii{\m{b}_{0}}_{\m{G}}$
    \For{$i = 1$ to $k$}\Comment{$\m{G}$-Lanczos process}
        \State $\m{w}_{i} \gets \tilde{\m{A}} \m{G} \m{u}_{i}$
        \State $\alpha_{i} \gets \innpr{\m{u}_{i}}{\m{w}_{i}}_{\m{G}}$
        \State $\m{r}_{i} \gets \m{w}_{i} - \alpha_{i} \m{u}_{i} - \beta_{i-1} \m{u}_{i-1}$\Comment{Residual vector}
        \State $\beta_{i} \gets \vertii{\m{r}_{i}}_{\m{G}}$
        \State $\m{u}_{i} \gets \tilde{\m{w}}_{i} / \beta_{i}$
        \State 
    \EndFor
    \State $\m{U}_{k} \gets [\m{u}_{1} \vert \m{u}_{2} \vert \cdots \vert \m{u}_{k}], \, \m{T}_{k} \gets \text{Tridiagonal}([\beta, \alpha, \beta])$
    \State $([\lambda_{1}, \cdots, \lambda_{l}], [\m{s}_{1} \vert \cdots \vert \m{s}_{l}]) \gets \text{Eigen}(l, \m{T}_{k})$\Comment{Incomplete eigendecomposition of $\m{T}_{k}$}
    \State $[\m{v}_{1} \vert \cdots \vert \m{v}_{l}] \gets \m{U}_{k} [\m{s}_{1} \vert \cdots \vert \m{s}_{l}]$
\EndProcedure
\end{algorithmic}
\end{algorithm}

\section{Numerical Study}
This section provides the numerical results; additional details are provided in \cref{sec:add:simul}. Code for reproducing the results is available in the open-source \href{https://github.com/HoYUN-stat/CovIterSolvers.jl}{Julia package}, with accompanying \href{https://hoyun-stat.github.io/CovIterSolvers.jl/dev/}{documentation}.

All simulations consider zero-mean Gaussian processes over the interval $\mathcal{T} = [0, 1]$. This centering avoids discrepancies between the two estimation formulations discussed in \cref{rmk:centred}, ensuring that $\tilde{\m{A}} = \m{A}$. For the \texttt{TReK} method, to minimize tuning complexity, we apply early termination as a form of regularization \cite{hanke2001lanczos, engl1996regularization, hanke2017conjugate}, setting the Tikhonov parameter $\eta = 0$ in \cref{thm:dict:2nd:mom,thm:rkhs:2nd:mom} when solving
\begin{equation*}
    \min_{\tilde{\m{A}}} \|\s{O} (\m{y} \odot \m{y}) - \s{F} \tilde{\m{A}}\|^{2}.
\end{equation*}
Since the resulting linear system is often numerically inconsistent -- meaning $\s{O} (\m{y} \odot \m{y}) \notin \mathcal{R}(\s{F})$ -- we replace CGLS with LSQR for the dictionary methods and MINRES for the RKHS methods with a maximum iteration count of 50. In practice, convergence typically occurs well before reaching this limit, as illustrated in \cref{ssec:param:tune} and the \href{https://github.com/HoYUN-stat/CovIterSolvers.jl/tree/main/examples/01_animated_covariance_smoothing}{animation}.

To assess estimation accuracy, we use the relative Mean Integrated Squared Error (MISE):
\begin{equation*}
    \text{Rel.MISE}(\Sigma) := \frac{\|\hat{\Sigma} - \Sigma\|_{\c{L}_2(\c{T})}^{2}}{\|\Sigma\|_{\c{L}_2(\c{T})}^{2}} = \frac{\int_{0}^{1} \int_{0}^{1} (\hat{\Sigma}(s, t) - \Sigma(s, t))^{2} ds dt}{\int_{0}^{1} \int_{0}^{1} \Sigma(s, t)^{2} ds dt}
\end{equation*}
For the evaluation of Mercer FPCA in the dictionary setting, we report both eigenfunction and eigenvalue errors for the top three components:
\begin{align*}
    \mathrm{MISE}(f_{k}) := \|\hat{f}_{k} - f_{k} \|_{\c{L}_2(\c{T})}^{2}
    = \int_{0}^{1} (\hat{f}_{k}(t) - f_{k}(t))^{2} dt, \quad
    \text{Rel.MSE}(\lambda_{k}) := \frac{|\hat{\lambda}_{k} - \lambda_{k}|^{2}}{\lambda_{k}^{2}}, \quad k=1, 2, 3.
\end{align*}

This section is organized as follows. \cref{ssec:param:tune} discusses strategies for selecting key estimation parameters in \texttt{TReK}, including the iteration count and the choice of search spaces. \cref{ssec:num:eff} illustrates the practical advantages of element-free Krylov methods over direct solvers in terms of computational efficiency. Finally, \cref{ssec:num:acc} evaluates the estimation accuracy using the tuned parameters, benchmarking against \texttt{fdapace} and \texttt{fpca.sc} across various experimental setups.

\subsection{Parameter Tuning}\label{ssec:param:tune}
While the Morozov discrepancy principle \cite{morozov1966solution, kaipio2006statistical} is widely used to prevent overfitting in unregularized problems, it is not applicable here due to the aforementioned numerical inconsistency of the linear system. Instead, we employ the L-curve as the stopping criterion \cite{hansen1993use, lawson1995solving}, which seeks a \emph{corner} in the log-log plot of the solution norm versus the gradient-residual norm over iterations, analogous to the \emph{elbow} in a PCA scree plot. In our context, the solution norm of the covariance function $\Sigma$ (with matrix representation $\tilde{\m{A}}$) is given by $\tr (\m{G} \tilde{\m{A}} \m{G} \tilde{\m{A}})$ for dictionary methods and $\tr (\m{K} \tilde{\m{A}} \m{K} \tilde{\m{A}})$ for RKHS methods, where $\m{G}$ is the Galerkin matrix from \cref{thm:fpca:dict} and $\m{K}$ is the mean kernel Gram matrix from \cref{thm:fpca:rkhs}. 

\begin{figure}[h!]
\centering
\includegraphics[width=.8\textwidth]{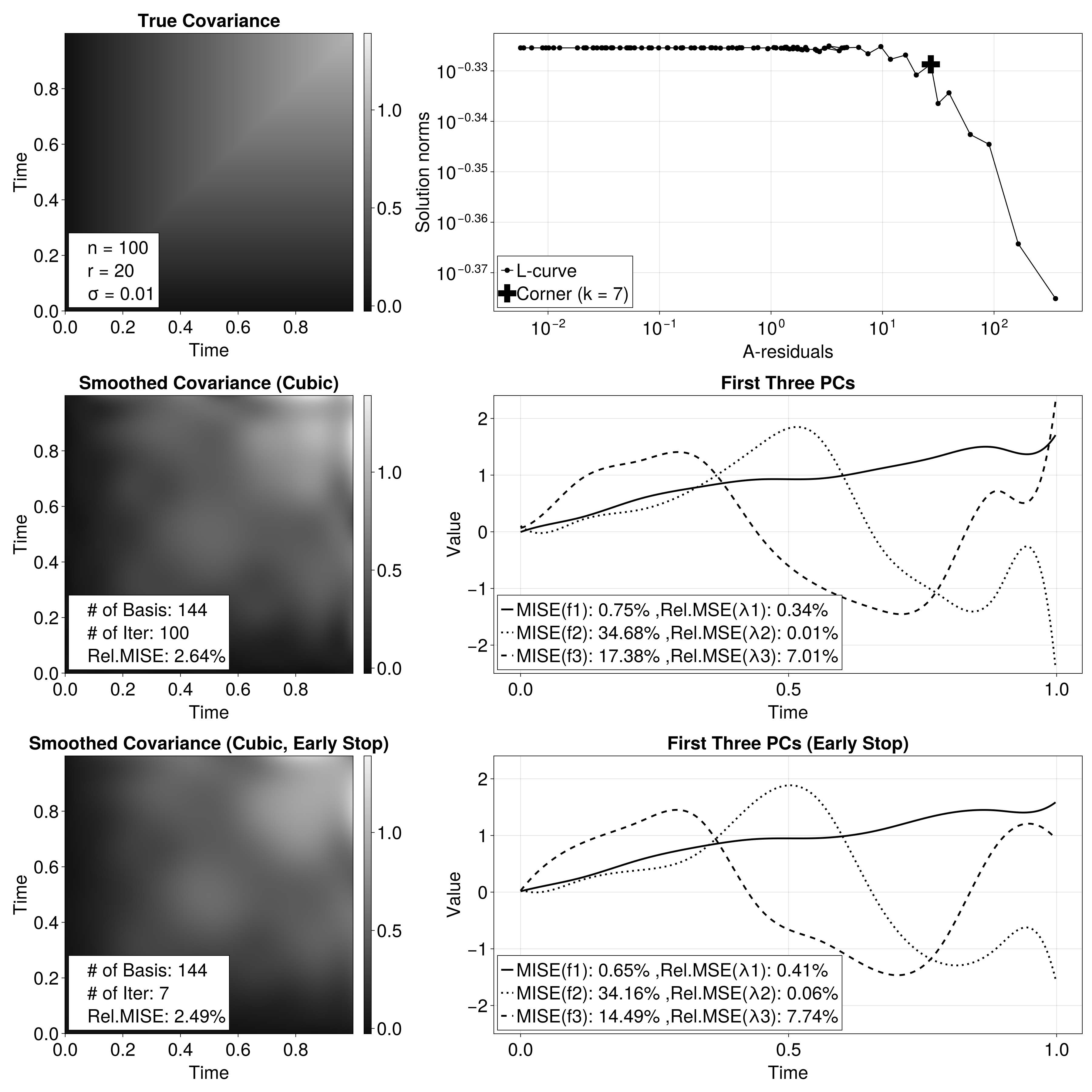}
\caption{Data was simulated from $n=100$ paths, each observed at $r=20$ points with noise $\sigma=0.01$.
\textbf{Top :} The true covariance structure of BM is shown for reference (left). The L-curve plots the trade-off between the solution and residual norms, identifying an optimal corner at $k=7$ iterations (right). 
\textit{Middle :} Running the estimation for a full 100 iterations results in overfitting. This produces a noisy covariance estimate and highly oscillatory Principal Components (PCs), especially at boundaries.
\textbf{Bottom :} Stopping at the L-curve corner ($k=7$) provides effective regularization, yielding a smooth covariance estimate and stable PCs. This demonstrates how early stopping can produce qualitatively superior functions, even in the covariance surface's relative MISE.}
\label{fig:earlystop}
\end{figure}

\cref{fig:earlystop} demonstrates this process using sample paths generated from Brownian Motion (BM), whose eigenvales and eigenfunctions of the Mercer expansion are given by:
\begin{equation*} 
    \lambda_{k} = [(k-1/2)\pi]^{-2}, \, f_{k}(t) = \sqrt{2} \sin \left( (k-1/2)\pi t \right), \quad k \in \b{N}, \, t \in \c{T}=[0, 1].
\end{equation*}
The figure illustrates the phenomenon of semi-convergence, where initial iterations reduce reconstruction error, but later iterations begin to overfit the noise in the data \cite{chung2024computational}.

The choice of function space for smoothing is another critical hyperparameter. To compare the performance of different bases, we generate $n=100$ sample paths from a Brownian Bridge (BB) process, whose covariance surface is more sharply concentrated along the diagonal than that of Brownian Motion, with the number of observations per path drawn from $r_{i} \stackrel{iid}{\sim} \mathrm{Unif}(10:12)$, and with observation noise $\sigma = 0.1$. We then apply the L-curve early stopping criterion to estimates derived from three types of bases, each with two hyperparameter configurations:
\begin{itemize}
    \item Cubic B-splines: $p=10$ and $p=20$.
    \item Gaussian Kernel: $K(s, t) = \exp (-\gamma (s-t)^2)$ with $\gamma = 2$ and $\gamma = 10$.
    \item Laplacian Kernel : $K(s, t) = \exp (-\gamma |s-t|)$ with $\gamma = 1$ and $\gamma = 3$.    
\end{itemize}

\begin{figure}[h!]
\centering
\includegraphics[width=\textwidth]{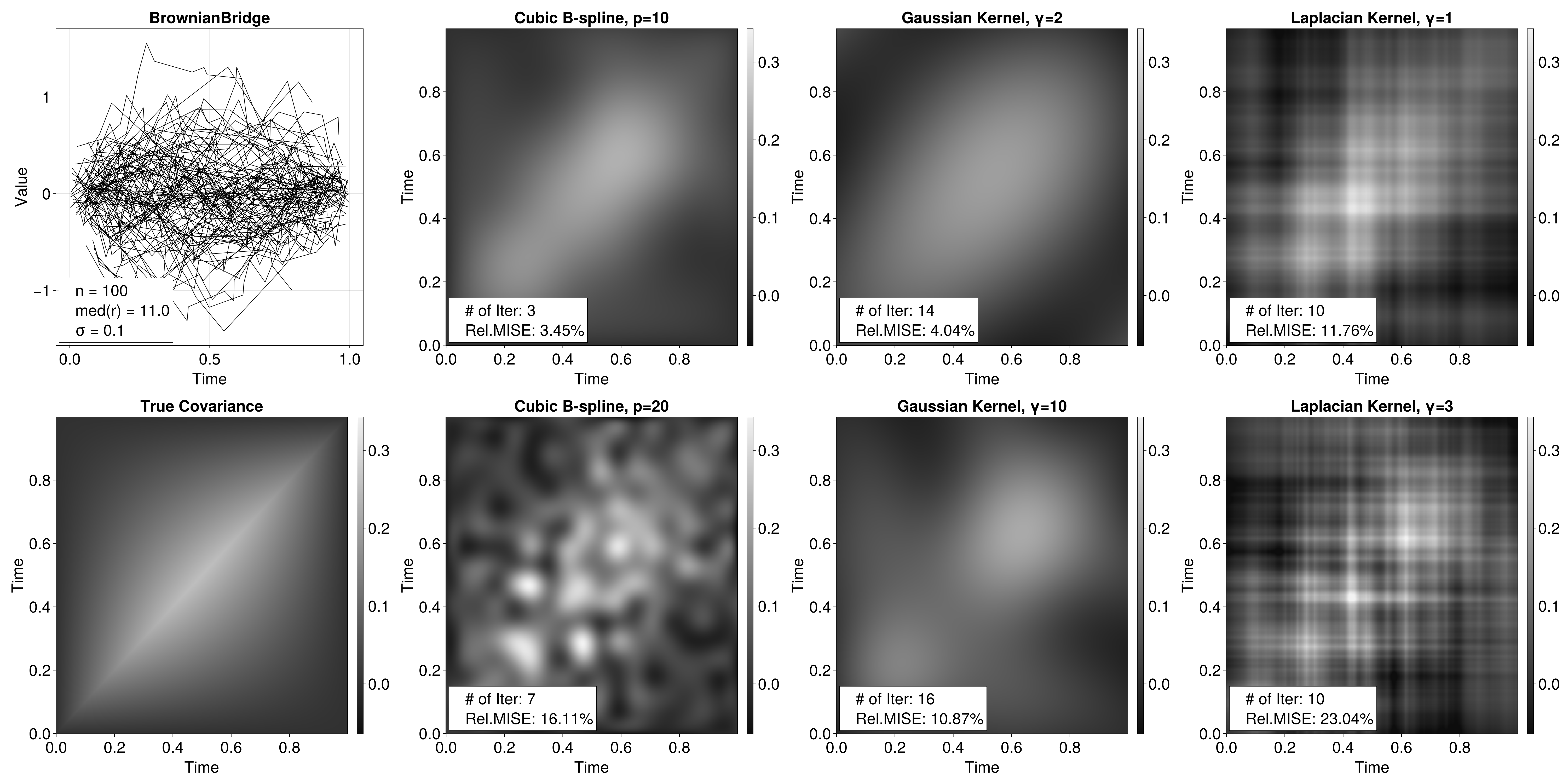}
\caption{Comparison of different smoothing bases for the covariance estimation of a BB process, with regularization by L-curve early stopping. The leftmost column shows the sample paths (top) and true covariance (bottom). The subsequent columns compare results from Cubic B-splines, Gaussian RKHS, and Laplacian RKHS with varying hyperparameters. The B-spline basis with $p=10$ achieves the lowest relative MISE (3.45\%) with the fewest iterations ($k=3$).}
\label{fig:basis:choice}
\end{figure}

The results, presented in \cref{fig:basis:choice}, indicate that the cubic B-spline basis with a smaller number of basis ($p=10$) is the most effective choice for \emph{smoothing} task of 1D longitudinal data structure. This approach is not only computationally less complex than the RKHS methods but also converges in significantly fewer iterations while achieving a lower relative MISE.
Furthermore, the performance of the RKHS methods proved highly sensitive to the locality parameter $\gamma$, with estimation quality degrading as $\gamma$ increases. 
This sensitivity is notable, as RKHS methods in other applications, such as computerized tomography, often require large $\gamma$ values to capture fine local details \cite{yun2025computerized, yun2025low}. Based on these findings, we exclusively use cubic B-splines with $p=10$ in the subsequent analyses of \cref{ssec:num:eff,ssec:num:acc}.  In practical scenarios where the true covariance is unknown, hyperparameters like $p$ or $\gamma$ can be effectively chosen using ($k$-fold) cross-validation \cite{cai2010nonparametric, caponera2022functional}. 

\subsection{Numerical Efficiency}\label{ssec:num:eff}
\begin{figure}[h!]
\centering
\includegraphics[width=\textwidth]{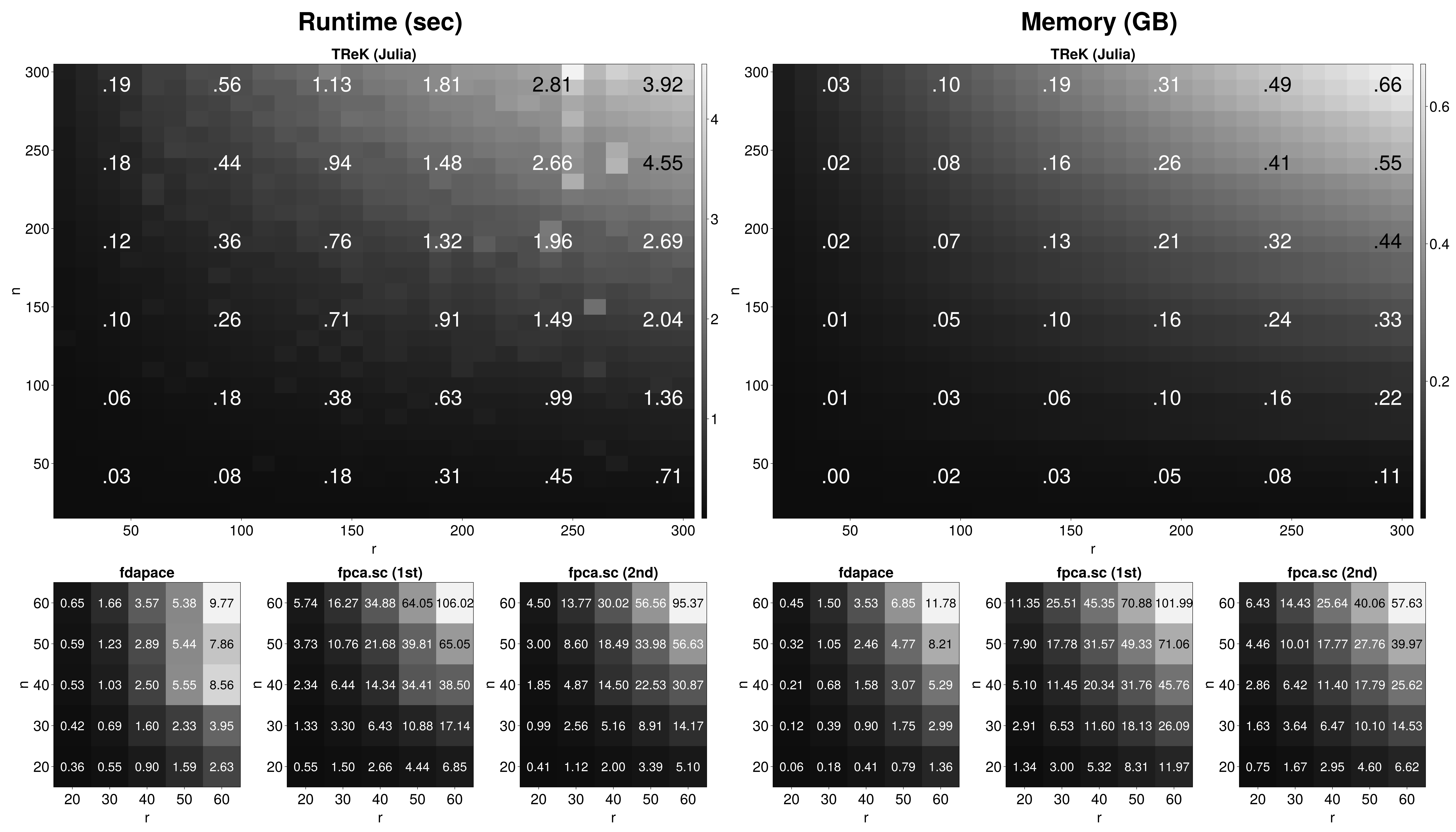}
\caption{Runtime (in seconds) and Memory (in gigabytes) performance benchmarks comparing \texttt{TReK} (Julia) with \texttt{fdapace} (R) and two estimation methods from \texttt{fpca.sc} (R) on an Apple M1 Pro. The heatmaps display median runtime (top) and memory allocation (bottom) across five repetitions. Data were generated from $n$ i.i.d. Brownian motion sample paths on the interval $\c{T} = [0, 1]$ with additive noise ($\sigma = 0.01$), each observed at $r$ random locations. Due to the high computational cost, the R packages were tested on a reduced range of $(n, r)$. Cubic B-splines with $p = 10$ were used for both \texttt{TReK} and \texttt{fpca.sc}, while \texttt{fdapace} used its default grid size ($g = 51$).
}
\label{fig:benchmark}
\end{figure}

We benchmark the numerical efficiency of the proposed \texttt{TReK} algorithm against two widely used R implementations: \texttt{fdapace} and \texttt{fpca.sc}. 
While precise comparisons across different high-level programming languages are inherently complex (e.g., compiler or garbage collection in Julia vs. R) , the results in \cref{fig:benchmark} suffice to reveal that \texttt{TReK} achieves superior scalability, reducing both runtime and memory usage by orders of magnitude in our tests, regardless of the configuration of $(n, r)$.
A fair comparison with \texttt{fpca.sc} is particularly relevant, as \texttt{TReK}'s B-spline approach solves a linear system of the same size as the second estimation method in \texttt{fpca.sc}, since both employ a ``pool-then-smooth'' strategy. To ensure a conservative benchmark that favors the competing methods, \texttt{TReK} was run for a fixed 50 iterations, even though it practically converges in substantially fewer, as demonstrated in \cref{ssec:param:tune}. It should be noted that the memory usage of \texttt{TReK} is independent of the iteration count, and \cref{fig:benchmark} shows that the memory usage adheres to our theoretical growth of space complexity proportional to $nr^{2}$. 

This efficiency has direct practical implications. As shown in \cref{fig:BM:Intro}, \texttt{TReK} successfully processes a large dataset of $n = 1000$ curves with $r_i \equiv r = 400$ observations each on a standard laptop, completing estimation in a few seconds. Although this scenario is extremely dense within the domain $\c{T} = [0, 1]$, it would be considered sparse in high-dimensional domains, e.g. $\c{T} = [0, 1]^{3}$. 
This scalability is particularly advantageous for high-dimensional applications such as fMRI \cite{sarkar2022covnet}. For these problems, the computational complexity can be reduced even further by leveraging the outer product structure in \cref{ssec:block:outer} across dimensions. A full exploration of such high-dimensional applications is beyond the scope of this paper and remains a direction for future work.

\subsection{Numerical Accuracy}\label{ssec:num:acc}
We evaluate the accuracy of \texttt{TReK} through a Monte Carlo simulation study. We consider four distinct data-generating scenarios by crossing two sizes of sample paths ($n \in \{100, 400\}$) with two noise levels ($\sigma \in \{0.01, 0.1\}$). For each scenario, we compare the performance of two estimation strategies: the fully iterated solution (50 iterations) and a regularized solution obtained via early stopping based on the L-curve criterion. The ground-truth covariance structure for all simulations is generated from the following Mercer expansion:
\begin{align*}
    \Sigma(s, t) = \sum_{k=1}^{3} \lambda_{k} f_{k}(s) f_{k}(t), \quad \lambda_{k} = 2^{1-k}, \, 
    f_{k}(t) = 
    \begin{cases}
        \sqrt{2} \sin(\pi t) &, \quad k=1, \\
        \sqrt{2} \cos(2 \pi t) &, \quad k=2, \\
        \sqrt{2} \sin(3 \pi t) &, \quad k=3.
    \end{cases}
\end{align*}
In each of the 200 Monte Carlo repetitions per setting, the number of observation points for each curve is drawn independently from $r_{i} \stackrel{iid}{\sim} \mathrm{Unif}(5, 15)$. Our \texttt{TReK} estimator employs a cubic B-spline basis with $p=10$ coefficients, defined by 8 equidistant knots on the domain $\mathcal{T} = [0, 1]$. The entire simulation, comprising $4 \times 2 \times 200 = 1600$ distinct estimations, completes in under one minute on 6 threads of Apple M1 Pro. 

\begin{figure}[h!]
\centering
\includegraphics[width=.8\textwidth]{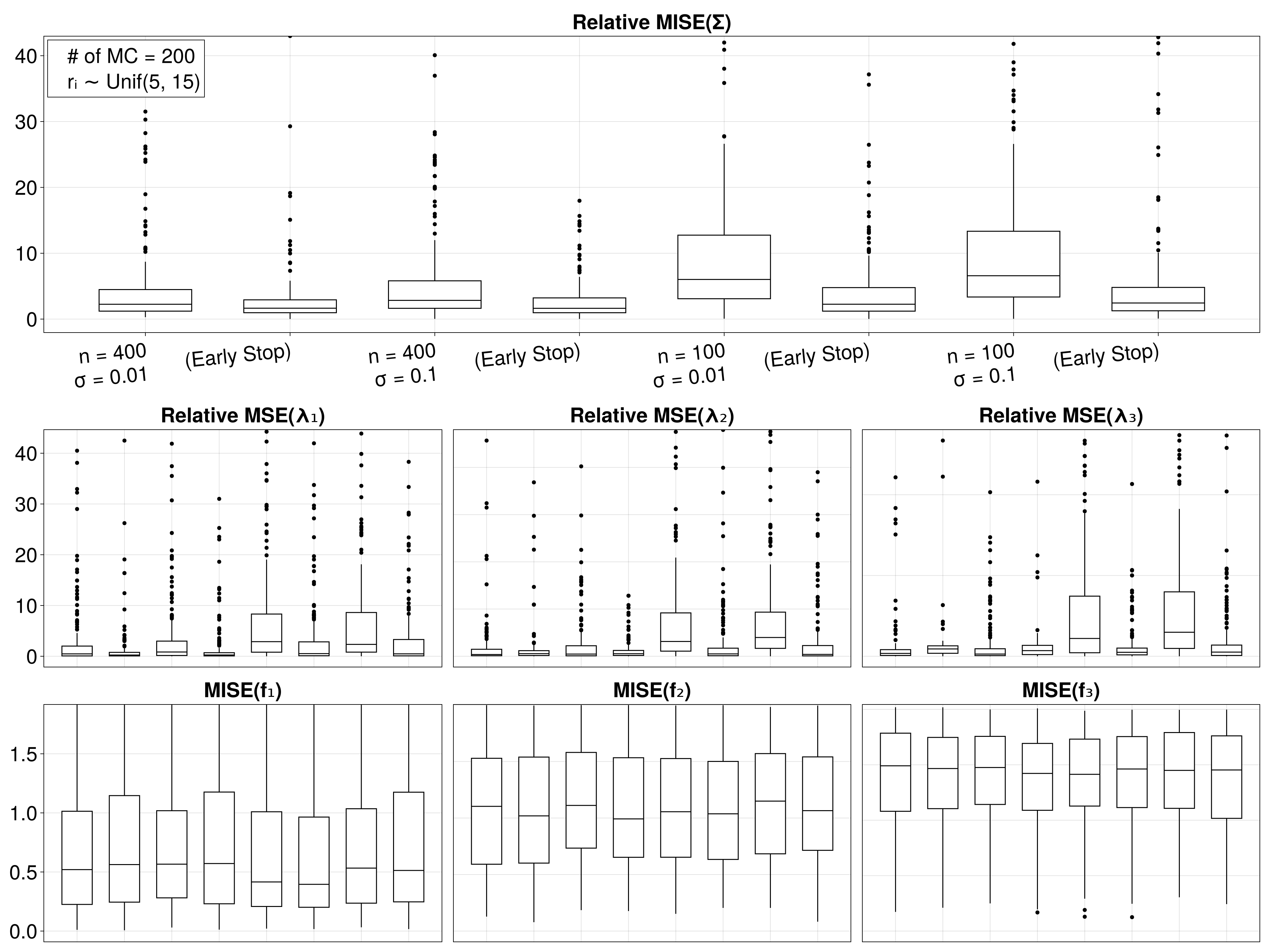}
\caption{Results of the Monte Carlo simulation study based on 200 repetitions per setting. The top row displays the relative MISE for the estimated covariance. The middle and bottom rows show the relative MSE for the first three eigenvalues ($\lambda_k$) and the MISE for their corresponding eigenfunctions ($f_k$), respectively. The x-axis categories are identical to those in the top panel and are omitted in the lower rows for visual clarity. Within each panel, boxplots compare the performance of the fully iterated solution against the one regularized by early stopping across the four data-generating scenarios (combinations of $n \in \{100, 400\}$ and $\sigma \in \{0.01, 0.1\}$).}
\label{fig:monte:carlo}
\end{figure}

The results are summarized in \cref{fig:monte:carlo}, which clearly demonstrate the effectiveness of early stopping as a regularization strategy; it consistently reduces estimation error and variance across all metrics, with the most significant gains in the high-noise ($\sigma=0.1$) scenarios. As expected, performance improves with a larger sample size ($n=400$) and lower noise level ($\sigma=0.01$).

\section{Conclusion}
The central message of this paper is to emphasize how abstraction in Hilbert space theory should inform and guide numerical methods in FDA. Unlike multivariate statistics -- where a standard orthonormal basis is readily available -- functional settings lack such a canonical coordinate system, and concepts like vectorization become contrived and ill-suited. This is precisely the aim of FPCA: to extract an intrinsic basis that reflects the geometry of the covariance operator. The coefficient matrix serves as a finite-dimensional proxy for the projection of this operator onto a chosen function subspace. Just as a matrix is fully characterized by its action on vectors, the same principle applies to operators acting on functions.
Krylov-based methods exploit this perspective, forming approximations using only repeated forward actions, without requiring full construction \cite{kammerer1972convergence}. This shift not only adheres to a genuinely functional perspective but also leads to substantial practical benefits.

Despite the centrality of linear operators in FDA, Krylov subspace methods and their variants remain underutilized in this context. This paper demonstrates how these methods provide an economical and efficient means of estimating covariance structures. By bridging operator-theoretic insights with numerical strategies, this work offers a principled and scalable framework for FPCA -- and opens the door to broader applications of Krylov-based methods in FDA.

\begin{acks}[Acknowledgments]
We thank the associate editor and the three referees for their insightful comments and constructive suggestions, which have significantly improved the manuscript. We are also grateful to Almond St\"{o}cker and Jake Grainger. A. St\"{o}cker provided valuable insights into both the practical and theoretical challenges of covariance smoothing. J. Grainger assisted with code verification, and shared  many practical tips that greatly improved the efficiency of the package over its preliminary version.
\end{acks}

\bibliographystyle{imsart-number}
\bibliography{bibliography}

\begin{appendix}
\section{Computational Complexity}\label{sec:cplxty}
We analyze the computational complexity of the operations involved. The time or space complexity of multiplying two square matrices of size $r \times r$ is denoted as $O(r^{2+\Delta})$, where $\Delta > 0$ depends on the specific algorithm used. For a classical approach, Strassen's algorithm \cite{strassen1969gaussian}, and the state-of-the-art approaches \cite{le2014algebraic,williams2024new} achieve $\Delta = 1$, $\log_{2}7-2 \approx 0.807$, and $\approx 0.376$, respectively. However, in our setting, the matrix multiplications are generally block-wise and not necessarily square. For the typical block dimensions encountered in covariance smoothing, these advanced algorithms often underperform due to their intricate data layout requirements and larger constant factors. Therefore, we adopt the standard assumption that multiplying an $a \times b$ matrix with a $b \times c$ matrix incurs a computational cost of $O(abc)$. The following proposition shows that the main computational cost of the second-order forward action is governed by the partitioned outer products rather than the elimination operator, as listed in \cref{table:LSE:2nd} in the Introduction.

\begin{proposition}\label{prop:cplxty}
The computational complexity of the elimination operator $\s{O}$, the partitioned outer products $\m{\Phi} \odot \m{\Phi}$ (or $\m{\Phi}^{*} \odot \m{\Phi}^{*}$), and $\m{K} \odot \m{K}$ are given by 
$O(|\m{r}|)$, $O(|\m{r}| p^{2} + |\m{r}^{2}|p)$, and $O(|\m{r}| |\m{r}^{2}|)$, respectively.
\end{proposition}
\begin{proof}[Proof of \cref{prop:cplxty}]
It is trivial that the complexity of $\s{O}$ is $\sum_{i=1}^{n} O(r_{i}) = O(|\m{r}|)$. Since $\m{\Phi}_{i} \in \b{R}^{r_{i} \times p}$, the complexity of both $(\m{\Phi}_{i} \otimes \m{\Phi}_{i})$ and $(\m{\Phi}_{i}^{\top} \otimes \m{\Phi}_{i}^{\top})$ is given by $O(r_{i}p^{2} + r_{i}^{2}p)$, hence the complexity of both $\m{\Phi} \odot \m{\Phi}$ and $\m{\Phi}^{*} \odot \m{\Phi}^{*}$ is given by
\begin{equation*}
    \sum_{i=1}^{n} O(r_{i}p^{2} + r_{i}^{2}p) = O(|\m{r}| p^{2} + |\m{r}^{2}|p).
\end{equation*}
Finally, for each $1 \le i \le n$, the complexity of $(\m{K}_{i\cdot} \otimes \m{K}_{i\cdot})$ is given by $\sum_{i'=1}^{n} O(r_{i}r_{i'}^{2} + r_{i}^{2}r_{i'})$,
hence the complexity of $\m{K} \odot \m{K}$ is given by
\begin{align*}
    O \left( \sum_{i, i'=1}^{n} (r_{i}r_{i'}^{2} + r_{i}^{2}r_{i'}) \right) = O \left( \sum_{i, i'=1}^{n} r_{i}r_{i'}^{2} \right) = O(|\m{r}| |\m{r}^{2}|).
\end{align*}
\end{proof}

In all of the outlined methods, a core computation is the symmetric sandwich product $\m{D}_{i} = \m{F}_{i} \m{C}_{i} \m{F}_{i}^{\top}$, across all blocks. Since the matrix block $\m{C}_{i}$ is symmetric (of size $p \times p$ for dictionary methods, and $r_{i} \times r_{i}$ for RKHS methods), which guarantees that the resulting block $\m{D}_{i}$ (of size $r_{i} \times r_{i}$) is also symmetric. Theoretically, this known symmetry could be exploited by specialized routines to reduce floating-point operations (FLOPs). 
However, modern numerical libraries in \texttt{Julia}, \texttt{R}, and \texttt{Python}, delegate matrix operations to highly optimized Basic Linear Algebra Subprograms (\texttt{BLAS}), the \emph{de facto} standard low-level API for such tasks. As outlined in \cref{alg:gem:sandwich}, the standard approach uses two general matrix-matrix multiplications (via the \texttt{GEMM} routine). While one might expect to accelerate this algorithm by signaling the symmetry to invoke a specialized routine \texttt{SYMM}, we empricially witnessed that, for the block dimensions typical in covariance smoothing, this provides no discernible speedup and can often be slower. Arguably, this discrepancy stems from the fact that, the usual \texttt{GEMM} routine is the most heavily optimized computational kernel, engineered to maximize memory throughput and exploit instruction-level parallelism (\texttt{SIMD}) on modern CPUs \cite{goto2008anatomy}. These hardware-specific optimizations often outweigh the theoretical FLOP reduction offered by \texttt{SYMM}.

\begin{algorithm}[h!]
\caption{GEMM Sandwich}\label{alg:gem:sandwich}
\begin{algorithmic}[1]
\Require $\m{F} \in \b{R}^{r_{i} \times p}, \m{C}_{i} \in \b{R}^{p \times p}$
\Ensure $\m{D}_{i} = \m{F}_{i} \m{C}_{i} \m{F}_{i}^{\top} \in \b{R}^{r_{i} \times r_{i}}$
    \State $\m{T}_{i} \gets \m{F}_{i} \m{C}_{i}$\Comment{Temporary buffer}
    \State $\m{D}_{i} \gets \m{T}_{i} \m{F}_{i}^{\top}$
\end{algorithmic}
\end{algorithm}

Therefore, we should decouple the meaning of \emph{fast} and \emph{cheap} in existing implementation of covariance smoothers when solving the matrix-vector form of \eqref{eq:dict:2nd:normal}:
\begin{align*}
    (\s{F}^{*} \s{F} + \lambda \s{D}^{*} \s{D}) \hat{\m{A}}(\lambda) = \s{F}^{*} (\m{y} \odot \m{y}).
\end{align*}
In their direct methods, exploiting symmetry via half-vectorization in the preprocessing step, where they exxplicitly construct the forward matrix, is primarily a memory-saving strategy (\emph{cheap}), which reduces the storage by a factor of $\approx 2^{2}$. This smaller memory footprint then makes the subsequent solver step \emph{faster} by a factor of $\approx 2^{3}$. However, it is important to note that underlying tensor structure is lost in the solving step.

On the contrary, \texttt{TReK} method takes the \emph{operator-based} approach, hence significantly \emph{cheaper} by an order of magnitude to above methods. Instead of storing an operator $\s{F}$ (of size $|\m{r}^{2}| \times p^{2}$ for dictionary methods, or $|\m{r}^{2}| \times |\m{r}^{2}|$ for RKHS methods), it only requires storing its components and a small, reusable buffer for computing operator-matrix products (\cref{alg:gem:sandwich}) on-the-fly. This approach preserves the underlying tensor structure, making the action of $\s{F}$ highly efficient, as listed in \cref{table:LSE:2nd}. As detailed in Table \ref{table:memory:cost}, the memory savings are substantial compared to direct methods, requiring only minimal reusable workspace for the operator action and the Krylov solver \cite{saad2003iterative, paige1982algorithm, greenbaum1997iterative}.


\begin{table}[h!]
\renewcommand{\arraystretch}{1.5}
\caption{Workspace (memory) requirements for Krylov-based estimation. The element-free approach avoids storing the full system operator, leading to significant memory savings. For further memory reduction, $\m{F}$ and $\m{y} \odot \m{y}$ could also be stored implicitly, while these compounds are not the main computational costs of an algorithm.}
\centering
\begin{tabular}{|c||c|c|}
\hline
Method & Dictionary-based & RKHS-based \\
\hline
\hline
Storing Data Matrix $\m{y} \odot \m{y}$ &  $|\m{r}^{2}|$ &  $|\m{r}^{2}|$  \\
\hline
Workspace for Operator $\s{F}$ & $\max(r_{i}) \times p$ & $\max(r_{i}) \times \max(r_{i})$  \\
\hline
Workspace for Krylov Solver & $4|\m{r}^{2}| + 2 p^{2}$ (LSQR) & $6|\m{r}^{2}|$ (MINRES)  \\
\hline
\end{tabular}
\label{table:memory:cost}
\end{table}

\section{Technical Proofs}

\begin{proof}[Proof of \cref{thm:dict:2nd:mom}]
Note that
\begin{align*}
    \Gamma(t_{ij_{1}}, t_{ij_{2}}) &= \sum_{l_{1}, l_{2} = 1}^{p} a_{l_{1} l_{2}} \phi_{l_{1}}(t_{ij_{1}}) \phi_{l_{2}}(t_{ij_{2}}) \\
    &= \sum_{l_{1}, l_{2} = 1}^{p} \m{\Phi}_{i}[j_{1}, l_{1}] \m{A}[l_{1}, l_{2}] \m{\Phi}_{i}[j_{2}, l_{2}]
    = (\m{\Phi}_{i} \m{A} \m{\Phi}_{i}^{\top})[j_{1}, j_{2}],
\end{align*}
which results in
\begin{align*}
    \c{L}_{\otimes}(\m{A}) &= \sum_{i=1}^{n} \sum_{(j_{1}, j_{2}) \in \c{O}_{i}} \left(y_{ij_{1}} y_{ij_{2}} -  \Gamma(t_{ij_{1}}, t_{ij_{2}}) \right)^{2} \\
    &=  \sum_{i=1}^{n} \vertii{\s{O}_{i}[\m{y}_{i} \m{y}_{i}^{\top} - (\m{\Phi}_{i} \otimes \m{\Phi}_{i}) \m{A}]}^{2}
    = \vertii{\s{O} (\m{y} \odot \m{y}) - \s{O} (\m{\Phi} \odot \m{\Phi}) \m{A}]}^{2}.
\end{align*}
Since $\s{O}$ is an orthogonal projection, this can be decomposed into
\begin{align*}
    \c{L}_{\otimes}(\m{A})
    &= \vertii{(\m{y} \odot \m{y}) - \s{O} (\m{\Phi} \odot \m{\Phi}) \m{A}]}^{2} - \vertii{(\s{I} - \s{O}) (\m{y} \odot \m{y})}^{2} \\
    &= \vertii{(\m{y} \odot \m{y}) - \s{F} \m{A}]}^{2} - \vertii{(\s{I} - \s{O}) (\m{y} \odot \m{y})}^{2},
\end{align*}
and the G\^{a}teaux derivative \cite{hsing2015theoretical} becomes $D \c{L}_{\otimes}(\m{A}) = 2 \s{F}^{*} [\s{F} \m{A}- (\m{y} \odot \m{y})]$. Also, the G\^{a}teaux derivative of the penalty term is $2 \s{P} \m{A}$, leading to the normal equation  in \eqref{eq:dict:2nd:normal}.
\end{proof}

\begin{proof}[Proof of \cref{thm:rkhs:2nd:mom}]
Note that
\begin{align*}
    \Gamma(t_{ij_{1}}, t_{ij_{2}}) &= \innpr{\Gamma}{\rk_{ij_{1}} \otimes \rk_{ij_{2}}} =
    \sum_{i'=1}^{n} \sum_{j'_{1}, j'_{2} =1}^{r_{i'}} a_{i'j'_{1}j'_{2}} \innpr{\rk_{i'j'_{1}} \otimes \rk_{i'j'_{2}}}{\rk_{ij_{1}} \otimes \rk_{ij_{2}}} \\
    &= \sum_{i'=1}^{n} \sum_{j'_{1}, j'_{2} =1}^{r_{i'}} a_{i'j'_{1}j'_{2}} \innpr{\rk_{i'j'_{1}}}{\rk_{ij_{1}}} \innpr{\rk_{i'j'_{2}}}{\rk_{ij_{2}}} \\
    &= \sum_{i'=1}^{n} \sum_{j'_{1}, j'_{2} =1}^{r_{i'}} \m{K}_{ii'}[j_{1}, j'_{1}] \m{A}_{i'}[j'_{1}, j'_{2}] \m{K}_{i'i}[j'_{2}, j_{2}] \\
    &= \sum_{i'=1}^{n} (\m{K}_{ii'} \m{A}_{i'} \m{K}_{i'i}) [j'_{2}, j_{2}] =[(\m{K}_{i\cdot} \otimes \m{K}_{i\cdot}) \m{A}] [j'_{2}, j_{2}].
\end{align*}
This leads to
\begin{align*}
    \c{L}_{\otimes}(\m{A}) &= \sum_{i=1}^{n} \sum_{(j_{1}, j_{2}) \in \c{O}_{i}} \left(y_{ij_{1}} y_{ij_{2}} -  \Gamma(t_{ij_{1}}, t_{ij_{2}}) \right)^{2} \\
    &=  \sum_{i=1}^{n} \vertii{\s{O}_{i}[\m{y}_{i} \m{y}_{i}^{\top} - (\m{K}_{i\cdot} \otimes \m{K}_{i\cdot}) \m{A}]}^{2}
    = \vertii{\s{O} (\m{y} \odot \m{y}) - \s{O} (\m{K} \odot \m{K}) \m{A}]}^{2}.
\end{align*}
Since $\m{A} \in \c{R}(\s{O})$ and $\s{O}$ is an orthogonal projection, this can be formed into
\begin{align*}
    \c{L}_{\otimes}(\m{A})
    &= \vertii{(\m{y} \odot \m{y}) - \s{O} (\m{K} \odot \m{K}) \s{O} \m{A}]}^{2} - \vertii{(\s{I} - \s{O}) (\m{y} \odot \m{y})}^{2} \\
    &= \vertii{(\m{y} \odot \m{y}) - \s{F} \m{A}]}^{2} - \vertii{(\s{I} - \s{O}) (\m{y} \odot \m{y})}^{2},
\end{align*}
and the rest of the proof is equivalent to \cref{thm:dict:2nd:mom}.
\end{proof}

We omit the proof of \cref{thm:fpca:dict} as it is same to that of below:
\begin{proof}[Proof of \cref{thm:fpca:rkhs}]
First, we show the $\m{K}$-orthogonality:
\begin{align*}
    \delta_{k_{1}k_{2}} = \innpr{f_{k_{1}}}{f_{k_{2}}}_{\b{H}} &= \innpr{\sum_{i_{1}=1}^{n} \sum_{j_{1}=1}^{r_{i_{1}}} v_{i_{1}j_{1}, k_{1}} \rk_{i_{1}j_{1}}}{\sum_{i_{2}=1}^{n} \sum_{j_{2}=1}^{r_{i_{2}}} v_{i_{2}j_{2}, k_{2}} \rk_{i_{2}j_{2}}}_{\b{H}} \\
    &= \sum_{i_{1}j_{1}} \sum_{i_{2}j_{2}} v_{i_{1}j_{1}, k_{1}} \innpr{\rk_{i_{1}j_{1}}}{\rk_{i_{2}j_{2}}}_{\b{H}} v_{i_{2}j_{2}, k_{2}} \\
    &= \sum_{i_{1}j_{1}} \sum_{i_{2}j_{2}} \m{V}^{*}[k_{1}, i_{1}j_{1}] \m{K}[i_{1}j_{1}, i_{2}j_{2}] \m{V}[i_{2}j_{2}, k_{2}] = [\m{V}^{*} \m{K} \m{V}] [k_{1}, k_{2}].
\end{align*}
Next, observe that
\begin{align*}
    \lambda_{k} f_{k} &= \Sigma f_{k} = \left( \sum_{i_{1}j_{1}} \sum_{i_{2}j_{2}} \tilde{a}_{i_{1} j_{1}, i_{2} j_{2}} \rk_{i_{1} j_{1}} \otimes_{\b{H}} \rk_{i_{2} j_{2}} \right) \sum_{i_{3}j_{3}} v_{i_{3}j_{3}, k} \rk_{i_{3}j_{3}} \\
    &= \sum_{i_{1}j_{1}} \sum_{i_{2}j_{2}} \sum_{i_{3}j_{3}} \tilde{a}_{i_{1} j_{1}, i_{2} j_{2}} \innpr{\rk_{i_{2}j_{2}}}{\rk_{i_{3}j_{3}}}_{\b{H}} v_{i_{3}j_{3}, k} \rk_{i_{1} j_{1}} \\
    &= \sum_{i_{1}j_{1}} \left( \sum_{i_{2}j_{2}} \sum_{i_{3}j_{3}} \tilde{\m{A}}[i_{1} j_{1}, i_{2} j_{2}] \m{K}[i_{2}j_{2}, i_{3}j_{3}] \m{V}[i_{3}j_{3}, k] \right) \rk_{i_{1} j_{1}} 
    = \sum_{i_{1}j_{1}} [\tilde{\m{A}} \m{K} \m{V}][i_{1}j_{1}, k] \rk_{i_{1} j_{1}},
\end{align*}
which leads to
\begin{align}\label{eq:sq:root:eigen}
    &0 = \Sigma f_{k} - \lambda_{k} f_{k} = \sum_{ij} [\tilde{\m{A}} \m{K} \m{V} - \m{V} \m{\Lambda}][ij, k] \rk_{i_{1} j_{1}} = 0, \quad 1 \le k \le |\m{r}| \nonumber \\
    &\Longleftrightarrow \quad [\tilde{\m{A}} \m{K} \m{V} - \m{V} \m{\Lambda}]^{\top} \m{K} [\tilde{\m{A}} \m{K} \m{V} - \m{V} \m{\Lambda}] = \m{0} \nonumber \\
    &\Longleftrightarrow \quad [\m{K}^{1/2} \tilde{\m{A}} \m{K}^{1/2}] (\m{K}^{1/2} \m{v}_{k}) = \lambda_{k} (\m{K}^{1/2} \m{v}_{k}), \quad 1 \le k \le |\m{r}|.
\end{align}
Note that whenever $\m{K}^{1/2} \m{v}_{k} = \m{K}^{1/2} \tilde{\m{v}}_{k}$, we have 
\begin{equation*}
    \vertii{\sum_{ij} (v_{ij, k} - \tilde{v}_{ij, k}) \rk_{ij}}_{\b{H}}^{2} = (\m{v}_{k} - \tilde{\m{v}}_{k})^{\top} \m{K} (\m{v}_{k} - \tilde{\m{v}}_{k}) = 0,
\end{equation*}
which demonstrates that the FPCA does not depend on the specific choice of $\m{v}_{k}$ satisfying \eqref{eq:sq:root:eigen}. In this regard, for each $1 \le k \le |\m{r}|$, and choose any $\m{v}_{k}$ that satisfies \eqref{eq:sq:root:eigen} and define $\tilde{\m{v}}_{k} := \m{v}_{k} - \m{n}_{k}$ where
\begin{equation*}
    \lambda_{k} \m{n}_{k} = (\tilde{\m{A}} \m{K} \m{v}_{k} - \lambda_{k} \m{v}_{k}).
\end{equation*}
Then $\tilde{\m{v}}_{k}$ satisfies
\begin{equation*}
    \tilde{\m{A}} \m{K} \tilde{\m{v}}_{k} - \lambda_{k} \tilde{\m{v}}_{k} = (\tilde{\m{A}} \m{K} \m{v}_{k} - \lambda_{k} \m{v}_{k}) - \lambda_{k} \m{n}_{k} = 0,
\end{equation*}
and $\tilde{\m{V}} = [\m{v}_{1} \vert \dots \vert \m{v}_{|\m{r}|}]$ satisfies $\tilde{\m{A}} \m{K} \tilde{\m{V}} = \tilde{\m{V}} \m{\Lambda}$, which completes the proof.
\end{proof}

\section{Additional Details}\label{sec:add:simul}
The results presented in this paper can be reproduced using our \texttt{Julia} software package, which is publicly available on \href{https://github.com/HoYUN-stat/CovIterSolvers.jl}{\texttt{Github}}. A tutorial in the repository explains the package's functionality:
\begin{itemize}[leftmargin=*]
\item \textbf{(Data Generation)}: Enables simulation from Gaussian processes using both built-in and custom covariance functions.
\item \textbf{(Multiple Dispatch)}: Supports multiple data precisions (e.g., \texttt{Float32} or \texttt{Float64}) for multi-threading and scalability to larger datasets.
\item \textbf{(Estimation Methods)}: Supports B-splines of arbitrary order and knot configurations, as well as any built-in or user-defined reproducing kernels. We use LSQR and MINRES as their respective default Krylov solvers, but any other Krylov methods implemented in \texttt{Krylov.jl} is compatible with our package.
\end{itemize}
All results in this paper were generated using double-float precision (\texttt{Float64}).

\end{appendix}
\end{document}